\documentclass[12pt,preprint]{aastex}

 \usepackage{epstopdf}

 \slugcomment{Accepted Version 2}

\newcommand\pv{\mbox{$p_V$}}
\newcommand\irfactor{\mbox{$p_{IR}/p_{V}$}}

\begin{document}

 \DeclareGraphicsExtensions{.pdf,.gif,.jpg}

 \title{NEOWISE Studies of Spectrophotometrically Classified Asteroids: Preliminary Results}
\author{A. Mainzer\altaffilmark{1}, T. Grav\altaffilmark{2}, J. Masiero\altaffilmark{1}, J. Bauer\altaffilmark{1}$^{,}$\altaffilmark{3}, E. Hand\altaffilmark{1}, D. Tholen\altaffilmark{4}, R. S. McMillan\altaffilmark{5}, T. Spahr\altaffilmark{6}, R. M. Cutri\altaffilmark{3}, E. Wright\altaffilmark{7}, J. Watkins\altaffilmark{8}, W. Mo\altaffilmark{2}, C. Maleszewski\altaffilmark{5}}

\altaffiltext{1}{Jet Propulsion Laboratory, California Institute of Technology, Pasadena, CA 91109 USA}
\altaffiltext{2}{Johns Hopkins University, Baltimore, MD}
\altaffiltext{3}{Infrared Processing and Analysis Center, California Institute of Technology, Pasadena, CA 91125, USA}
\altaffiltext{4}{Institute for Astronomy, University of Hawaii, 2680 Woodlawn Drive, Honolulu, Hawaii 96822-1839 USA}
\altaffiltext{5}{Lunar and Planetary Laboratory, University of Arizona, 1629 East University Blvd., Kuiper Space Science Bldg. \#92, Tucson, AZ 85721-0092, USA}
\altaffiltext{6}{Minor Planet Center, Harvard-Smithsonian Center for Astrophysics, 60 Garden Street, Cambridge, MA 02138 USA}
\altaffiltext{7}{UCLA Astronomy, PO Box 91547, Los Angeles, CA 90095-1547 USA}
\altaffiltext{8}{Department of Earth and Space Sciences, UCLA, 595 Charles Young Drive East, Box 951567, Los Angeles, CA 90095 USA}

 \begin{abstract}
 The NEOWISE dataset offers the opportunity to study the variations in albedo for asteroid classification schemes based on visible and near-infrared observations for a large sample of minor planets.  We have determined the albedos for nearly 1900 asteroids classified by the Tholen, Bus and Bus-DeMeo taxonomic classification schemes.  We find that the S-complex spans a broad range of bright albedos, partially overlapping the low albedo C-complex at small sizes.  As expected, the X-complex covers a wide range of albedos.  The multi-wavelength infrared coverage provided by NEOWISE allows determination of the reflectivity at 3.4 and 4.6 $\mu$m relative to the visible albedo.  The direct computation of the reflectivity at 3.4 and 4.6 $\mu$m enables a new means of comparing the various taxonomic classes. Although C, B, D and T asteroids all have similarly low visible albedos, the D and T types can be distinguished from the C and B types by examining their relative reflectance at 3.4 and 4.6 $\mu$m. All of the albedo distributions are strongly affected by selection biases against small, low albedo objects, as all objects selected for taxonomic classification were chosen according to their visible light brightness. Due to these strong selection biases, we are unable to determine whether or not there are correlations between size, albedo and space weathering.  We argue that the current set of classified asteroids makes any such correlations difficult to verify.  A sample of taxonomically classified asteroids drawn without significant albedo bias is needed in order to perform such an analysis.
 \end{abstract}

 \section{Introduction}
Determining the true compositions of asteroids would significantly enhance our understanding of the conditions and processes that took place during the formation of the Solar System.  It is necessary to study asteroids directly as weathering and geological processes have tended to destroy the oldest materials on Earth and the other terrestrial planets.  The asteroids represent the fragmentary remnants of the rocky planetesimals that built these worlds, and asteroids in the Main Belt and Trojan clouds are likely to have remained in place for billions of years (subject to collisional processing) \citep{Gaffey1993M}.  Many attempts have been made to determine the minerological composition of asteroids by studying variations in their visible and near-infrared (VNIR) spectroscopy and photometry \citep{Bus, Tholen84, Tholen89, Zellner, Binzel04, DeMeo09}.  Efforts have been made to link asteroid spectra with those of meteorites \citep[e.g.][]{Thomas}.  However, as noted by \citet{Gaffey10} and \citet{Chapman04}, space weathering can complicate the linkages between observed asteroid spectra and meteorites.  In addition, VNIR spectroscopic and photometric samples of higher-albedo objects are generally more readily attainable, as these bodies are brighter as compared with low albedo bodies with similar heliocentric distances and sizes.  An important element in the development of asteroid taxonomic schemes has been albedo.  For example, in the classification system developed by \citet{Tholen84}, the E, M and P classes have degenerate Eight-Color Asteroid Survey \citep[ECAS;][]{Zellner} spectra and can only be distinguished by albedos.  All of these issues point to the need to a) obtain a large, uniform sample of asteroid albedos (and other physical properties such as thermal inertia) that can be compared with VNIR classifications, and b) expand the number of asteroids with VNIR classifications in order to bracket the full range of asteroid types and compositions.

With the \emph{Wide-field Infrared Survey Explorer's} NEOWISE project \citep{Wright, Mainzer11a}, thermal observations of more than 157,000 asteroids throughout the Solar System are now in hand, a dataset nearly two orders of magnitude larger than that provided by the \emph{Infrared Astronomical Satellite} \citep[IRAS;][]{Matson, Tedesco02}.  Thermal models have been applied to these data to derive albedos and diameters for which taxonomic classifications are available.  In this paper, we examine the NEOWISE-derived albedos and diameters for NEOs and Main Belt asteroids of various classification schemes based on visible and NIR spectroscopy and multiwavelength spectrophotometry.  In a future work, we will compare NEOWISE albedos to classifications and visible/NIR colors found photometrically, such as with the Sloan Digital Sky Survey or $BVR$ photometry.  The taxonomic classes and NEOWISE-derived albedos of the Trojan asteroids are discussed in \citet{Grav}. While many different asteroid classification schemes have been created, we turn our focus initially to three commonly used schemes, those defined by \citet{Tholen84}, \citet{Bus}, and \citet{DeMeo09}.    

\section{Observations}
WISE is a NASA Medium-class Explorer mission designed to survey the entire sky in four infrared wavelengths, 3.4, 4.6, 12 and 22 $\mu$m (denoted $W1$, $W2$, $W3$, and $W4$ respectively) \citep{Wright, Liu, Mainzer}.  The final mission data products are a multi-epoch image atlas and source catalogs that will serve as an important legacy for future research.   The survey has yielded observations of over 157,000 minor planets, including NEOs, MBAs, comets, Hildas, Trojans, Centaurs, and scattered disk objects \citep{Mainzer11a}.  

The observations for the objects listed in Table 1 were retrieved by querying the Minor Planet Center's (MPC) observation files to look for all instances of individual NEOWISE detections of the desired objects that were reported during the cryogenic portion of the mission using the WISE Moving Object Processing System \citep[WMOPS;][]{Mainzer11a}.  The data for each source were extracted from the WISE First Pass Processing archive following the methods described in \citet{Mainzer11b}.  The artifact identification flag cc\_flags (which specifies whether or not an instrumental artifact was likely to have occurred on top of a given source) was allowed to be equal to either 0, p, or P, and the flag ph\_qual (which describes whether the source was considered a valid detection) was restricted to A, B, or C \citep[a comprehensive explanation of these flags is given in][]{Cutri}.  As described in \citet{Mainzer11b}, we used observations with magnitudes close to experimentally-derived saturation limits, but when sources became brighter than $W1=6$, $W2=6$, $W3=4$ and $W4=0$, we increased the error bars on these points to 0.2 magnitudes and applied a linear correction to $W3$ \citep{Cutri}. 

Each object had to be observed a minimum of three times at SNR $>$5 in at least one WISE band, and to avoid having low-level unflagged artifacts and/or cosmic rays contaminating our thermal model fits, we required that observations in more than one band appear with SNR$>$5 at least 40\% of the number of observations found in the band with the largest number of observations (usually $W3$).  If the number of observations exceeds the 40\% threshold, \emph{all} of the detections in that band are used.  Although this strategy could possibly cause us to overestimate fluxes and colors, the fact that we use all available observations when the minimum number of observations with SNR$>$5 has been reached gives us some robustness against this. This problem was identified with IRAS; see http://irsa.ipac.caltech.edu/IRASdocs/exp.sup/ch12/A.html\#1 for details.  We recognize this potential issue and will revisit it in a future work, particularly when we have the results from final version of the WISE data processing pipeline in hand. Artifact flagging and instrumental calibration will be substantially improved with the final version of the WISE data processing pipeline, and we will re-examine the issue of low-SNR detections and non-detections when these products are available.

The WMOPS pipeline rejected inertially fixed objects in bands $W3$ and $W4$ before identifying moving objects; however, it did not reject stationary sources in bands $W1$ and $W2$.  To ensure that asteroid detections were less likely to be confused with stars and background galaxies, we cross-correlated the individual Level 1b detections with the WISE atlas and daily coadd catalogs.  Objects within 6.5 arcsec (equivalent to the WISE beam size at bands $W1$, $W2$ and $W3$) of the asteroid position which appeared in the coadded source lists at least twice and which appeared more than 30\% of the total number of coverage of a given area of sky were considered to be inertially fixed sources; these asteroid detections were considered contaminated and were not used for thermal fitting.   

In this paper, we consider only NEOs or Main Belt asteroids that were observed during the fully cryogenic portion of the NEOWISE mission.  Results from the NEOWISE Post-Cryogenic Mission will be discussed in a future work.  For a discussion of WISE colors and physical properties derived from NEOWISE data for the bulk population of NEOs, see \citet{Mainzer11d}.  \citet{Masiero} and \citet{Grav} give WISE colors and thermal fit results for the Main Belt asteroids and Trojan asteroids observed during the cryogenic portion of the mission, respectively.   

\section{Preliminary Thermal Modeling of NEOs}
We have created preliminary thermal models for each asteroid using the First-Pass Data Processing Pipeline (version 3.5) described above; these thermal models will be recomputed when the final data processing is completed.  As described in \citet{Mainzer11b}, we employ the spherical near-Earth asteroid thermal model (NEATM) of \citet{Harris}.  The NEATM model uses the so-called beaming parameter $\eta$ to account for cases intermediate between zero thermal inertia \citep[the Standard Thermal Model; ][]{Lebofsky_Spencer} and high thermal inertia \citep[the Fast Rotating Model; ][]{Lebofsky78,Veeder89,Lebofsky_Spencer}. In the STM, $\eta$ is set to 0.756 to match the occultation diameters of (1) Ceres and (2) Pallas, while in the FRM, $\eta$ is equal to $\pi$.  With NEATM, $\eta$ is a free parameter that can be fit when two or more infrared bands are available (or with only one infrared band if diameter or albedo are known \emph{a priori} as is the case for objects that have been imaged by visiting spacecraft or observed with radar). 

Each object was modeled as a set of triangular facets covering a spherical surface with a variable diameter \citep[c.f.][]{Kaasalainen}. Although many (if not most) asteroids are non-spherical, the WISE observations generally consisted of $\sim$10-12 observations per object uniformly distributed over $\sim$36 hours \citep{Wright, Mainzer11a}, so on average, a wide range of rotational phases were sampled.  Although this helps to average out the effects of a rotating non-spherical object, caution must be exercised when interpreting the meaning of an effective diameter in these cases.  All diameters given are considered effective diameters, where the assumed sphere has a volume close to that of the actual body observed.  Tests with non-spherical triaxial ellipsoid models show that even for objects with peak-to-peak brightness variations of $\sim$1 mag, the derived diameter is found to have a 1-$\sigma$ error bar of $\sim$20\% compared to the effective diameter of the ellipsoid, provided that the rotational period is more than the average sampling frequency of 3 hours and less than the average coverage of $\sim$1 day \citep{Grav}.

 Thermal models were computed for each WISE measurement, ensuring that the correct Sun-observer-object distances were used. The temperature for each facet was computed, and the \citet{Wright} color corrections were applied to each facet. In addition, we adjusted the $W3$ effective wavelength blueward by 4\% from 11.5608$\mu$m to 11.0984 $\mu$m, the $W4$ effective wavelength redward by 2.5\% from 22.0883 $\mu$m to 22.6405 $\mu$m, and we included the -8\% and +4\% offsets to the $W3$ and $W4$ magnitude zeropoints (respectively) due to the red-blue calibrator discrepancy reported by \citet{Wright}.   The emitted thermal flux for each facet was calculated using NEATM; nightside facets were assumed to contribute no flux.   For NEOs, bands $W1$ and $W2$ typically contain a mix of reflected sunlight and thermal emission. The flux from reflected sunlight was computed for each WISE band as described in \citet{Mainzer11b} using the IAU phase curve correction \citep{Bowell}.  Facets which were illuminated by reflected sunlight and visible to WISE were corrected with the \citet{Wright} color corrections appropriate for a G2V star.  In order to compute the fraction of the total luminosity due to reflected sunlight, it was necessary to determine the relative reflectivity in bands $W1$ and $W2$.  This step is discussed in greater detail below.
   
In general, absolute magnitudes ($H$) were taken from the MPC's orbital element files.  The assumed $H$ error was taken to be 0.3 magnitudes.  Updated $H$ magnitudes were taken from the Lightcurve Database of \citet{Warner} for about 2/3 of the asteroids that were detected by NEOWISE that are considered herein.  Emissivity, $\epsilon$, was assumed to be 0.9 for all wavelengths \citep[c.f.][]{Harris09}, and $G$ (the slope parameter of the magnitude-phase relationship) was set to 0.15$\pm$0.10 based on \citet{Tholen09} unless a direct measurement from \citet{Warner} or \citet{Pravec} was available.  Accurate determination of albedo is critically dependent on the accuracy of the $H$ and $G$ values used for each asteroid; the albedos determined with the NEOWISE data will only be as accurate as the $H$ and $G$ values used to compute them.  We describe some instances in which we suspect that the assumption of $G=0.15$ is inappropriate below.  These objects will benefit from improved measurements of $G$.
 
For objects with measurements in two or more WISE bands dominated by thermal emission, the beaming parameter $\eta$ was determined using a least squares minimization but was constrained to be less than the upper bound set by the FRM case ($\pi$). As described in \citet{Mainzer11c}, the median value of the NEOs that had fitted $\eta$ was 1.41$\pm$0.5, while the weighted mean value was 1.35.  The beaming parameter could not be fitted for NEOs that had only a single WISE thermal band; these objects were assigned $\eta=1.35\pm0.5$. For Main Belt asteroids, the median value of the objects with fitted $\eta$ was 1.00$\pm$0.20 as discussed in \citet{Masiero}.  For MBAs with observations in only a single WISE thermal band, $\eta$ was set equal to 1.00$\pm$0.20.  

Bands $W1$ and $W2$ consist of a mix of reflected sunlight and thermal emission for NEOs, and bands $W3$ and $W4$ consist almost entirely of thermal emission.  In order to properly model the fraction of total emission due to reflected sunlight in each band, it was necessary to determine the ratio of the infrared albedo $p_{IR}$ to the visible albedo $p_{V}$.  We make the simplifying assumption that the reflectivity is the same in both bands $W1$ and $W2$, such that $p_{IR}=p_{3.4}=p_{4.6}$;  the validity of this assumption is discussed below.  The geometric albedo $p_{V}$ is defined as the ratio of the brightness of an object observed at zero phase angle ($\alpha$) to that of a perfectly diffusing Lambertian disk of the same radius located at the same distance.  The Bond albedo ($A$) is related to the visible geometric albedo $p_{V}$ by $A\approx A_{V} = qp_{V}$, where $q$ is the phase integral and is defined such that $q=2\int \Phi(\alpha) sin(\alpha) d\alpha$. $\Phi$ is the phase curve, and $q=1$ for $\Phi=max(0,cos(\alpha))$. $G$ is the slope parameter that describes the shape of the phase curve in the $H-G$ model of \citet{Bowell} that describes the relationship between an asteroid's brightness and the solar phase angle.  For $G=0.15$, $q=0.384$.  We make the assumption that $p_{IR}$ obeys the same relationship, although it is possible it varies with wavelength, so what we denote here as $p_{IR}$ for convenience may not be exactly analogous to $p_{V}$.  We can derive $p_{IR}/p_{V}$ for the WISE objects that have a significant fraction ($\sim50\%$ or more) of reflected sunlight in bands $W1$ and $W2$ as well as observations in $W3$ or $W4$. As discussed in \citet{Mainzer11d}, for the NEOs for which $p_{IR}/p_{V}$ could not be fitted, we used $p_{IR}/p_{V}=1.6\pm1.0$; per \citet{Masiero}, we set $p_{IR}/p_{V}=1.5\pm0.5$ for Main Belt asteroids.  For the objects with fitted $p_{IR}/p_{V}$, we can begin to study how reflectivity changes at 3.4 and 4.6 $\mu$m, and this can be compared to taxonomic types.

Where available, we used previously measured diameters from radar, stellar occultations, or in situ spacecraft imaging and allowed the thermal model to fit only $p_{IR}/p_{V}$ when $W1$ or $W2$ was available.  For a more complete description of the methodology and the sources of the diameter measurements, see \citet{Mainzer11b}.
 
 As described in \citet{Mainzer11b} and \citet{Mainzer11c}, the minimum diameter error that can be achieved using WISE observations is $\sim10\%$, and the minimum albedo error is $\sim20\%$ of the value of the albedo for objects with more than one WISE thermal band for which $\eta$ can be fitted.  For objects with large amplitude lightcurves, poor $H$ or $G$ measurements, or poor signal to noise measurements in the WISE bands, the errors will be higher.
 
\subsection{High Albedo Objects}
We note that among the asteroids considered here, there are $\sim$20 that have $p_{V}>0.65$. Approximately 2/3 of these objects have large peak-to-peak $W3$ variations, indicating that they are likely to be highly elongated or even binary.  In these cases, a spherical model is not likely to produce a good fit; these objects should be modeled as non-spherical shapes.  Almost all of the extremely high albedo objects are known to be members of the Vesta family or Hungarias.  It is possible that for these objects, the standard value of $G=0.15$, i.e. a fixed $q$ of 0.393, is not appropriate.  \citet{Harris1988} and \citet{Harris1989} noted that E and V type asteroids can have slope values as high as $G\sim$0.5.  The assumption of $G=0.15$ for an object like this would cause an error in the computed $H$ for observations at 20$^{\circ}$ phase angle of $\sim$0.3 magnitudes; this would drive the albedo derived using such an $H$ value up by 0.3, for example.  Albedos larger than 0.65 should be considered suspect; only a direct measurement of $G$ (and therefore $H$) for these objects will improve the reliability of the albedo determination for these objects. These objects would greatly benefit from additional study and more ground-based follow-up to improve their $H$ and $G$ values.   

\section{Discussion}
We have examined the taxonomic classifications of near-Earth objects (NEOs) and Main Belt asteroids (MBAs) provided by a number of groups.  \citet{Chapman75} proposed a series of letter-based taxonomic classes: S for stony or silicate-rich objects, C for carbonaceous asteroids, and U for asteroids that did not fit either class neatly.  \citet{Tholen84} defined seven major classes (A, C, D, E, M, P, and S), along with three subclasses of the C-complex (B, F and G), the minor class T, and the single-member classes R, Q, and V based on ECAS \citep{Zellner}.  Objects in the E, M and P classes could only be separated by their albedos as they were spectrally degenerate in the ECAS system; together, they form the X class.  The Tholen classification scheme relied upon ECAS; ECAS used a photometer with filters ranging from 0.34 $\mu$m to 1.04 $\mu$m. The ultraviolet wavelengths used by ECAS became more difficult to obtain when CCDs became widely available.  Visible CCD spectroscopy of asteroids was undertaken, and subsequent revisions to the taxonomic systems were made that no longer relied upon ultraviolet wavelengths.  

The Small Main-belt Asteroid Spectroscopic Survey \citep{Xu} and its second phase \citep[SMASSII;][]{Bus,Burbine,BusThesis} has produced visible spectroscopy for nearly 3000 asteroids.  From this dataset, \citet{Bus} defined three major groupings similar to \citet{Tholen84} (the S-, C- and X-complexes) and split them into 26 classes depending on the presence or absence of particular spectral features or slopes in visible wavelengths. In the system of \citet{Bus}, albedo is not used, and the short wavelength definition of the taxonomy extends only to 0.44 microns. Thus, limitations arise in that, for example, C- and X-types can be difficult to distinguish without albedo and without measurements over UV wavelengths.  \citet{DeMeo09} and \citet{DeMeoThesis} extended the system of \citet{Bus} by using near-infrared spectral features as well as visible, creating a system of 24 taxonomic types.  Neither the \citet{Bus} nor \citet{DeMeo09} systems use albedo as a means of taxonomic classification.  

Taxonomic classification systems can provide some understanding of the compositional nature of asteroids, but they have limitations.  Reflected colors may in some cases reveal mineral absorption bands that provide diagnostic information on composition, but the appearance of these spectral features can be influenced by other materials with similar absorption features, material states, particle sizes, illumination angles, etc.  Furthermore, some bodies' spectra are generally featureless.  For all of these reasons, other physical parameters such as albedo become important for further interpreting composition.  We have used the classification data compiled in the Planetary Data System Small Body Node by \citet{Neese}, which aggregates taxonomic types for $\sim$2600 minor planets from various sources. Table 1 gives the average albedos that we have computed from the asteroids we have observed with NEOWISE for each of the various taxonomic classes in the Tholen, Bus and Bus-DeMeo schemes.  A discussion of the biases that must be considered when comparing the albedos between classes is given below.

\begin{figure}
\figurenum{2}
\includegraphics[width=6in]{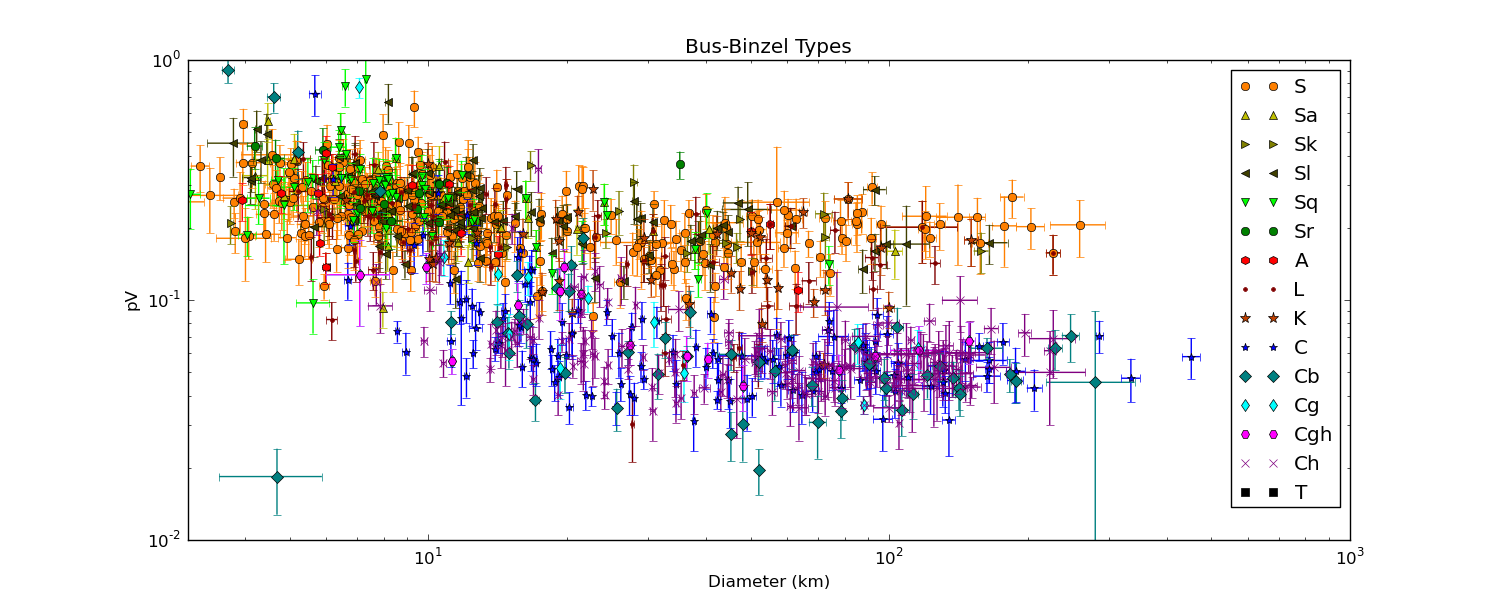}
\caption{\label{fig:binzel_diam_alb} NEOWISE-derived diameters vs. albedos of asteroids observed and classified according to the system of \citet{Bus}. The S and C complexes are shown; the X complex has been omitted for clarity.  There are few albedo distinctions evident among the subtypes in both the S and C complexes in the Bus-Binzel taxonomic system. As with the Tholen system shown in Figure 1, selection biases become apparent below $\sim$30 km and may be entirely responsible for the trend of increasing albedo with decreasing diameter.}
\end{figure}

\begin{figure}
\figurenum{3}
\includegraphics[width=6in]{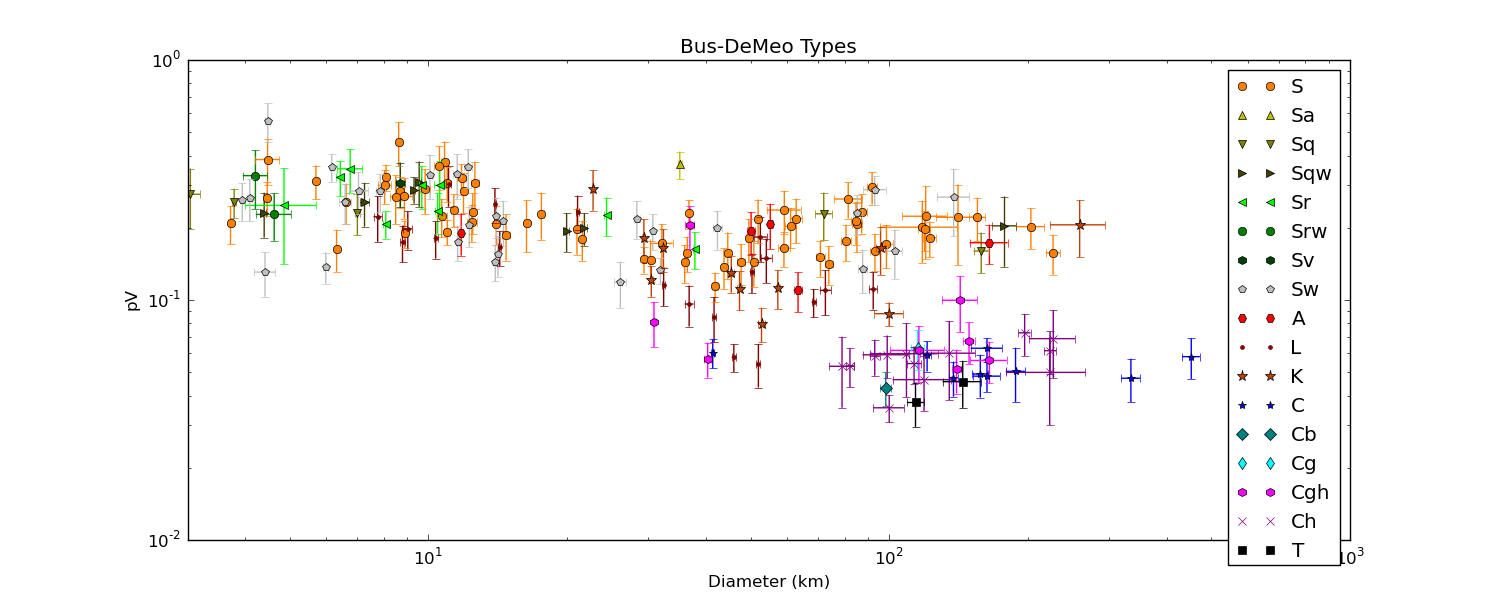}
\caption{\label{fig:demeo_diam_alb} NEOWISE-derived albedos of S and C complex asteroids observed and classified according to the taxonomic system of \citet{DeMeo09}, which supercedes the system of \citet{Bus}.  In this system, subtypes with a ``w'' have redder VNIR slopes and are supposed to be weathered versions of the original types; for example, Sw is the more reddened version of S.  However, no difference in albedo between the Sw and S types can be seen at all size ranges.  No differences among the C subtypes can be observed, although the comparison suffers from small number statistics.   }
\end{figure}

In Figure \ref{fig:tholen_diam_alb}, we show the diameter compared to \pv\ for 1247 asteroids observed and classified according to the Tholen scheme \citep{Tholen89,Xu,Lazzaro}, including 15 NEOs and 1232 MBAs.  Figure \ref{fig:binzel_diam_alb} shows diameter vs. \pv\ for the 1524 objects classified according to the Bus scheme \citep{Bus,Lazzaro}, including 21 NEOs and 1503 MBAs.  Finally, Figure \ref{fig:demeo_diam_alb} shows the 233 asteroids classified according to the DeMeo scheme (14 NEOs and 219 MBAs), which is based heavily on that of Bus. It should be noted that the same objects can have different classifications according to multiple schemes. Since so few NEOs have been observed relative to the numbers of MBAs, we have included the NEOs in our plots; there are not enough to significantly change the statistics. In all three schemes, an uptick in the average value of \pv\ for smaller diameters ($<$30 km) can be observed, regardless of spectral class.  There is a notable absence of small, dark objects, particularly among the C complex types, yet numerically low albedo objects represent the majority of the asteroids in the Main Belt \citep{Masiero}.   Although \citet{Delbo}, \citet{HarrisACM} and \citet{Wolters} have asserted that there is a real change in albedo with size, these studies are all based upon very small numbers of asteroids that are selected from visible light surveys. If there is a correlation between albedo and size, it is best studied using the full NEOWISE dataset rather than the relatively small population that has been selected from visible light surveys for spectroscopic study to date.  When we compare diameter to \pv\ for the entire NEOWISE set selected by the WMOPS pipeline \citep{Mainzer11a}, we find no strong trend of increasing \pv\ with decreasing diameter.  The selection bias in the population with taxonomic classifications acts twice. Objects with higher albedos are more likely to have been discovered by visible light surveys; a 5 km object with a 40\% albedo is nearly a full magnitude brighter than a 5 km object with a 20\% albedo. Similarly, the 40\% albedo object is more likely to have been selected for the spectroscopic studies necessary for taxonomic classification because it is more likely to be bright enough to observe.   We observe 11 out of 14 objects with a Bus-Binzel C complex classification with diameters between 6-10 km that have \pv$>$0.09, compared to 47 C types with diameters between 80 - 110 km with a median \pv$=$0.053$\pm$0.002 and standard deviation of 0.014.  However, these 11 small diameter outlier objects are entirely consistent with the number expected when we consider that the total population of low albedo small Main Belt asteroids numbers at least in the high tens of thousands \citep{Masiero}.  We cannot make reliable claims about possible relationships between size, albedo and space weathering without assembling a sample of asteroids in which the albedo biases are clearly understood.  As discussed below, what is needed is a spectroscopically classified sample that is unbiased with respect to albedo.  This study will be the subject of future work.    

\begin{figure}
\figurenum{4}
\includegraphics[width=6in]{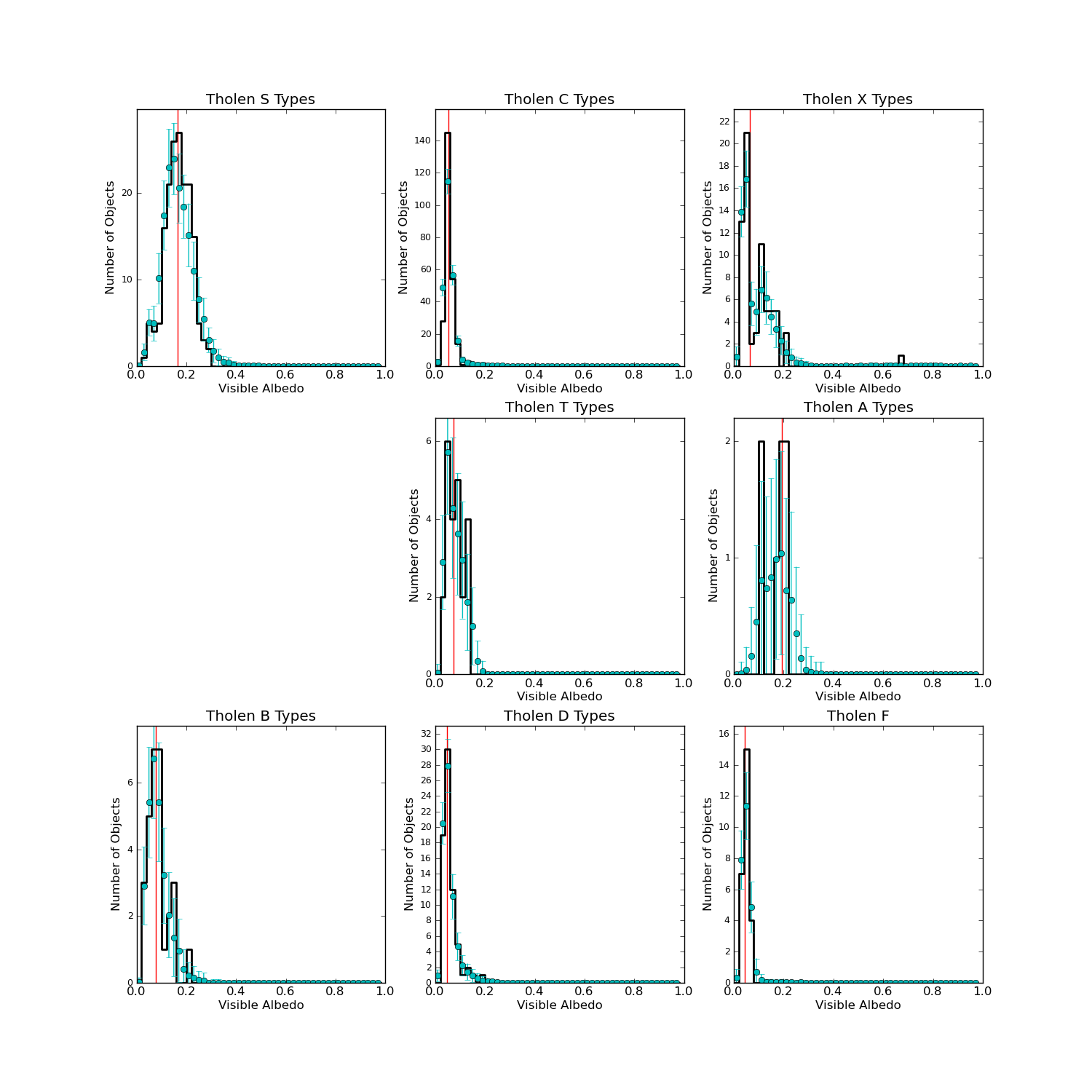}
\caption{\label{fig:TholenpV} NEOWISE-derived albedos of asteroids observed and classified by \citet{Tholen84} with diameters $>$ 30 km.  The dots with error bars represent the results of a 100 Monte Carlo simulation of the histogram using the error bars for each individual albedo measurement. The vertical red line represents the median \pv\ for each type. }
\end{figure}

\begin{figure}
\figurenum{5}
\includegraphics[width=6in]{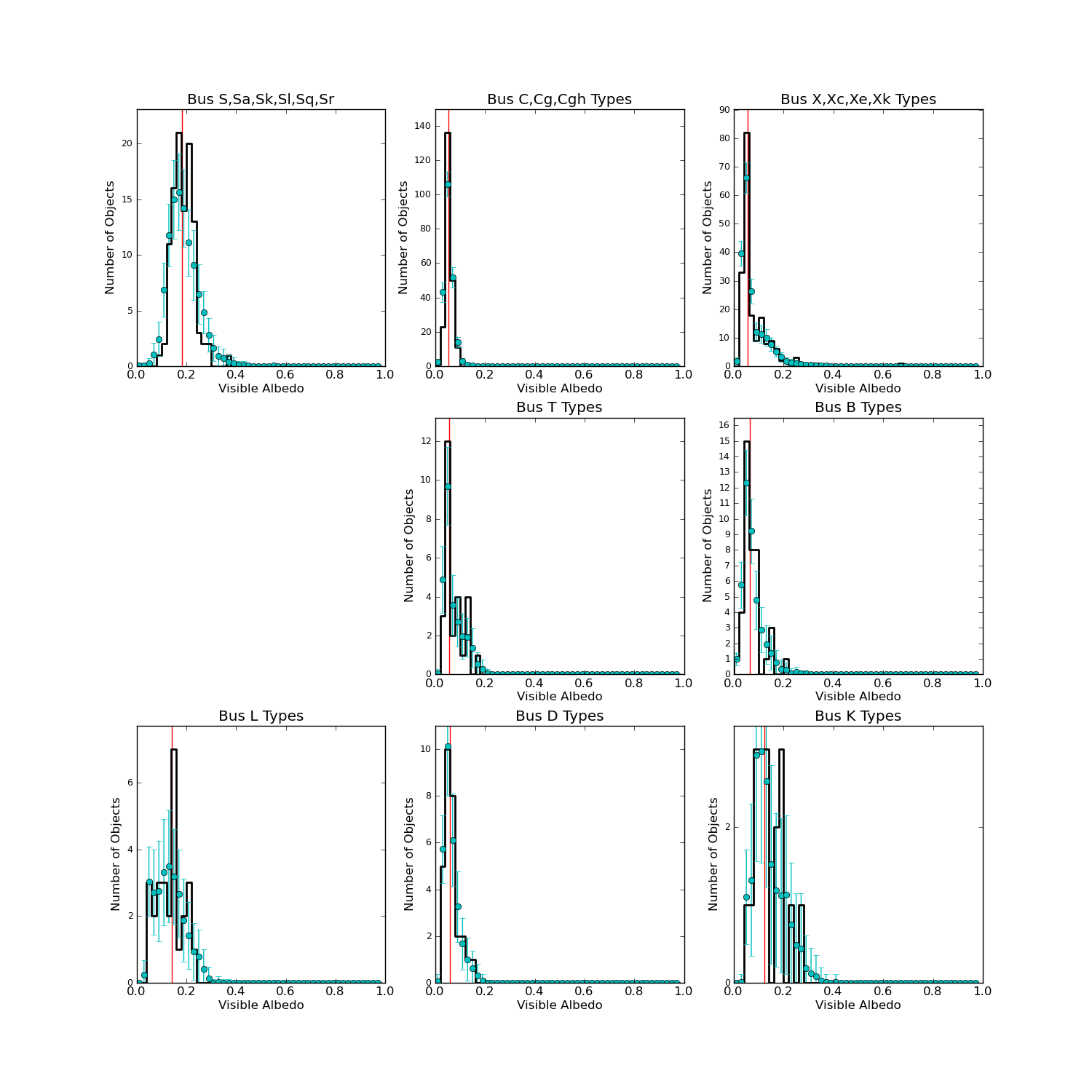}
\caption{\label{fig:BinzelpV} NEOWISE-derived albedos of asteroids observed and classified by \citet{Bus} with diameters $>$30 km.  The dots with error bars represent the results of a 100 Monte Carlo simulation of the histogram using the error bars for each individual albedo measurement. The vertical red line represents the median \pv\ for each type.}
\end{figure}

\begin{figure}
\figurenum{6}
\includegraphics[width=6in]{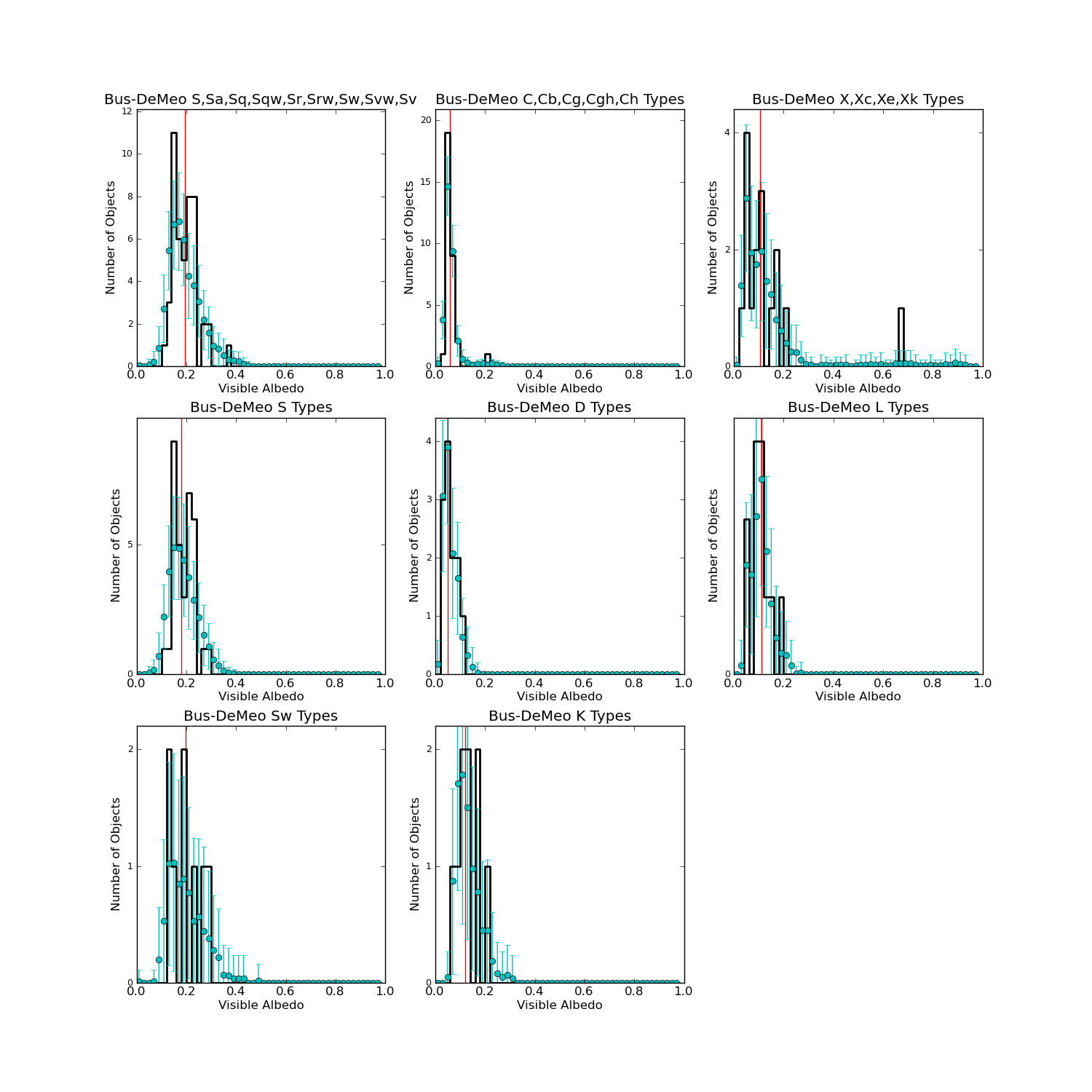}
\caption{\label{fig:BusDemeopV} NEOWISE-derived albedos of asteroids observed and classified by \citet{DeMeo09} with diameters $>$30 km.  The dots with error bars represent the results of a 100 Monte Carlo simulation of the histogram using the error bars for each individual albedo measurement. The vertical red line represents the median \pv\ for each type.}
\end{figure}

\begin{figure}
\figurenum{7}
\includegraphics[width=6in]{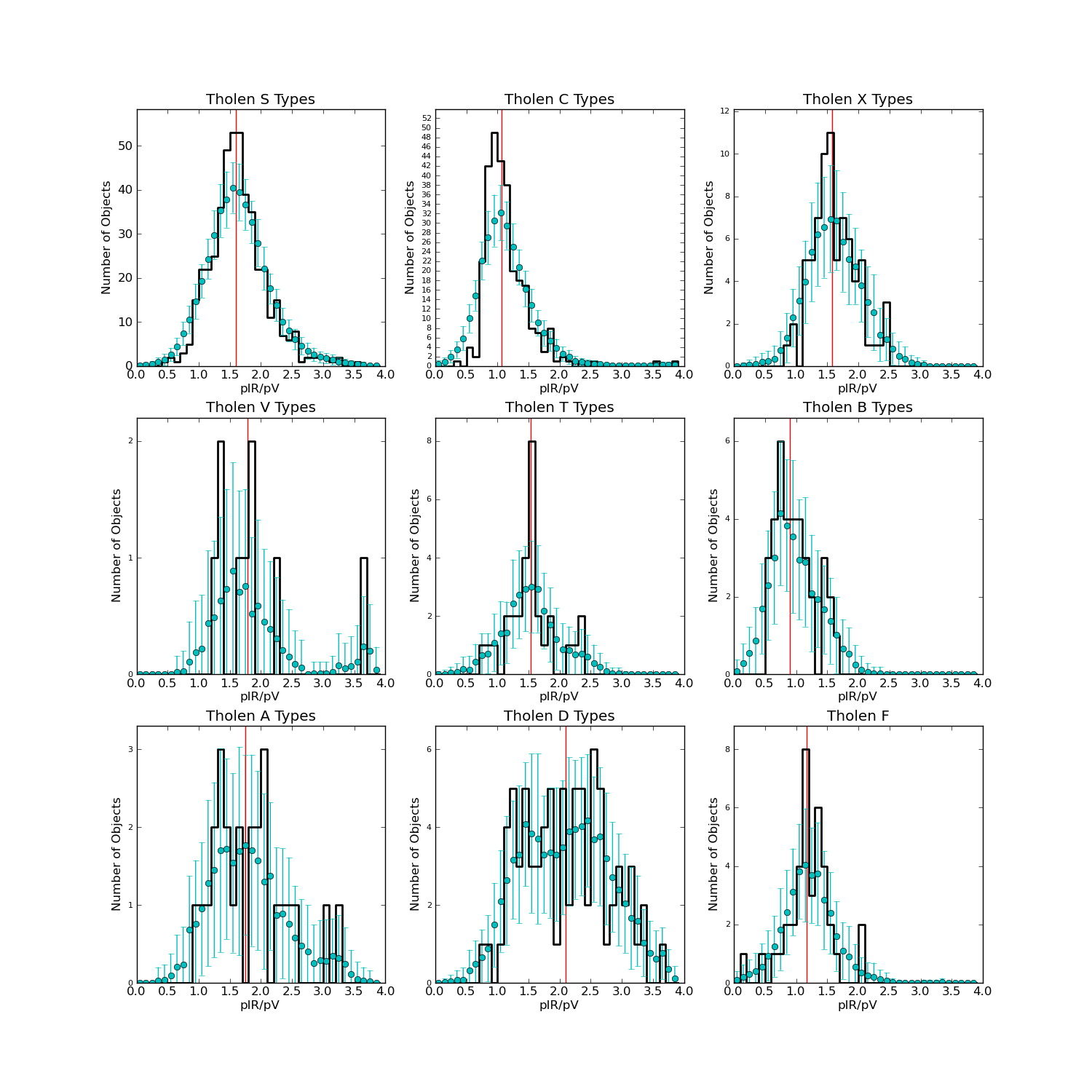}
\caption{\label{fig:TholenIR} NEOWISE-derived ratio $p_{IR}/p_{V}$ for asteroids observed and classified by \citet{Tholen84}.  Only asteroids for which $p_{IR}/p_{V}$ could be fitted are included in this plot. The dots with error bars represent the results of a 100 Monte Carlo simulation of the histogram using the error bars for each individual albedo measurement. The vertical red line represents the median \pv\ for each type.}
\end{figure}

\begin{figure}
\figurenum{8}
\includegraphics[width=6in]{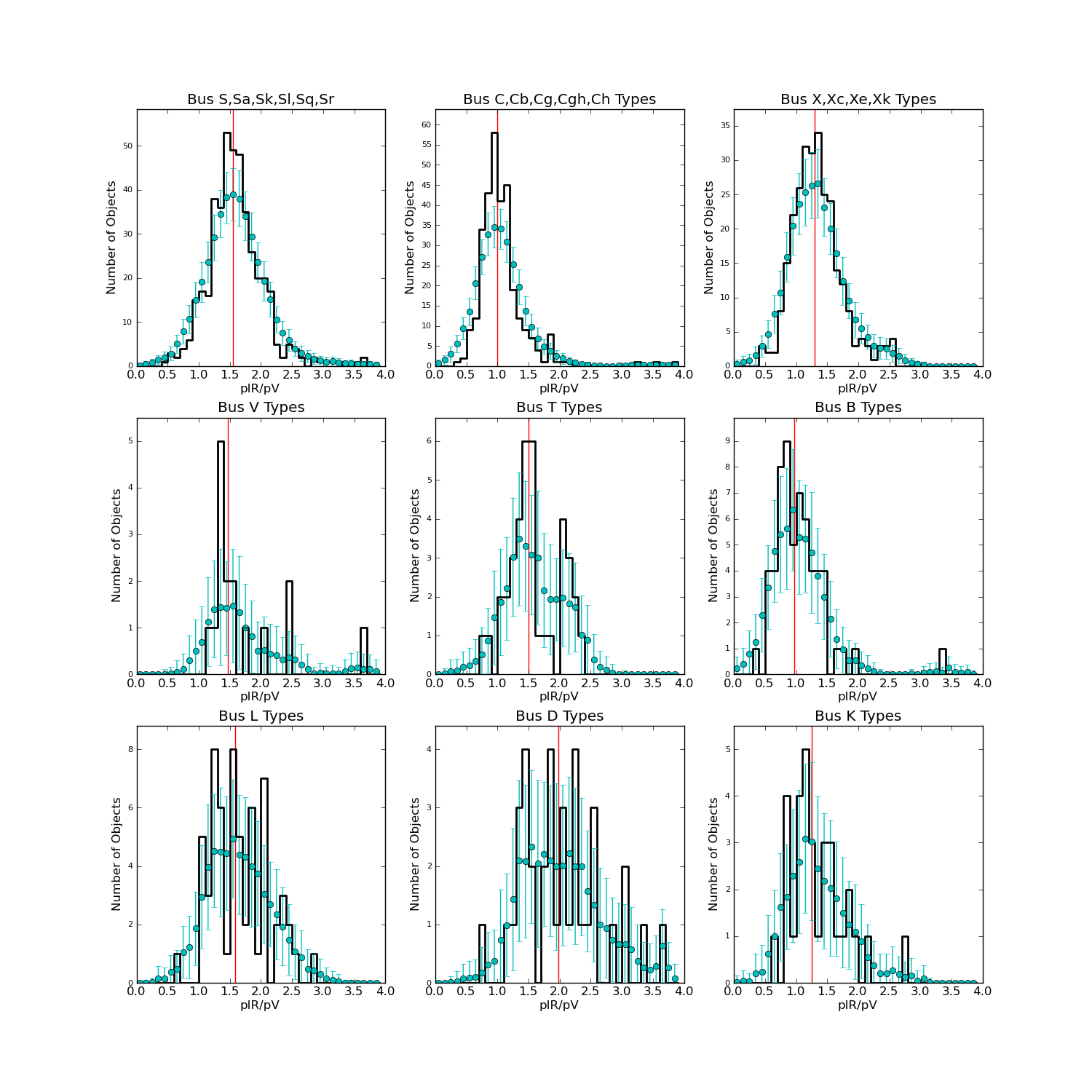}
\caption{\label{fig:BinzelIR} NEOWISE-derived ratio $p_{IR}/p_{V}$ for asteroids observed and classified by \citet{Bus}.  Only asteroids for which $p_{IR}/p_{V}$ could be fitted are included in this plot. The dots with error bars represent the results of a 100 Monte Carlo simulation of the histogram using the error bars for each individual albedo measurement.  The vertical red line represents the median \irfactor\ for each type.}
\end{figure}

\begin{figure}
\figurenum{9}
\includegraphics[width=6in]{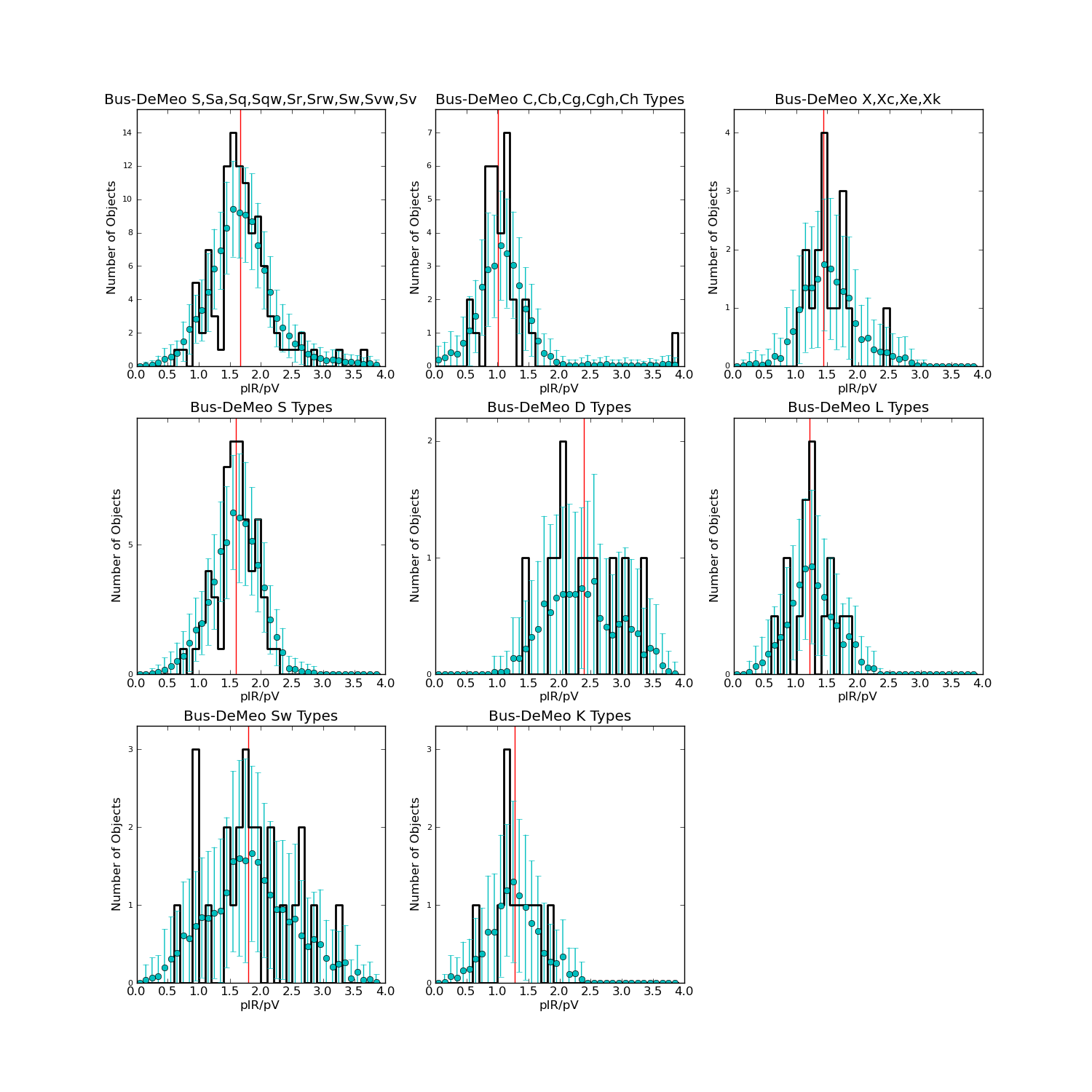}
\caption{\label{fig:BusDemeoIR} NEOWISE-derived ratio $p_{IR}/p_{V}$ for asteroids observed and classified by \citet{DeMeo09}.  Only asteroids for which $p_{IR}/p_{V}$ could be fitted are included in this plot. The objects have been separated broadly into S, C, X complexes with S, Sw, D and L types separated out since they each have more than a handful of objects.  The dots with error bars represent the results of a 100 Monte Carlo simulation of the histogram using the error bars for each individual albedo measurement.  The vertical red line represents the median \irfactor\ for each type.}
\end{figure}

From Figures \ref{fig:tholen_diam_alb}, \ref{fig:binzel_diam_alb} and \ref{fig:demeo_diam_alb}, we can see that there is generally good separation of \pv\ between S and C complex objects for diameters $>$30 km; we conclude that this is approximately the size down to which the visible light surveys are roughly complete.  In Figures \ref{fig:TholenpV}, \ref{fig:BinzelpV} and \ref{fig:BusDemeopV}, we show the visible albedo distributions for objects with diameters $>$30 km for the various spectral types in each of the three taxonomic systems.  In the Tholen system, 172 S type objects with diameters $>$30 km have a median \pv$=$0.166$\pm$0.004 with a standard deviation of 0.050, and 250 C type objects have a median \pv$=$0.053$\pm$0.002 with a standard deviation of 0.024.  In the system of \citet{Bus}, 106 S-complex objects with diameters $>$30 km (including S, Sa, Sk, Sl, Sq, and Sr) have a median \pv$=$0.182$\pm$0.004 with a standard deviation of 0.043, and 222 C-complex objects (including C, Cb, Cg, Cgh, and Ch) have a median \pv$=$0.053$\pm$0.001 with a standard deviation of 0.014.  As discussed in \citet{Mainzer11b}, the average albedo error is $\sim$20\% of the albedo value.  We suggest that those attempting to use spectral type as a proxy for \pv\ use these values when converting between $H$ and diameter, although as discussed above, it is unclear whether these values are still appropriate for objects at sizes smaller than $\sim$30 km.  Figures \ref{fig:binzel_diam_alb} and \ref{fig:demeo_diam_alb} show that little distinction can be observed between the various subtypes in the S and C complexes in the Bus and Bus-DeMeo schemes at all size ranges.  The albedo differences between various spectral types are best preserved in the system of \citet{Tholen84}.  Figures \ref{fig:TholenIR}, \ref{fig:BinzelIR} and \ref{fig:BusDemeoIR} give the ratio of the reflectivity in bands $W1$ and $W2$ compared with \pv\ for the Bus, Bus-DeMeo and Tholen schemes, respectively.  The mean, standard deviation of the mean, standard deviation, and minimum/maximum values of \pv\ and \irfactor\ for each class (including objects at all size ranges) are given in Table 1.  

\begin{figure}
\figurenum{10}
\includegraphics[width=6in]{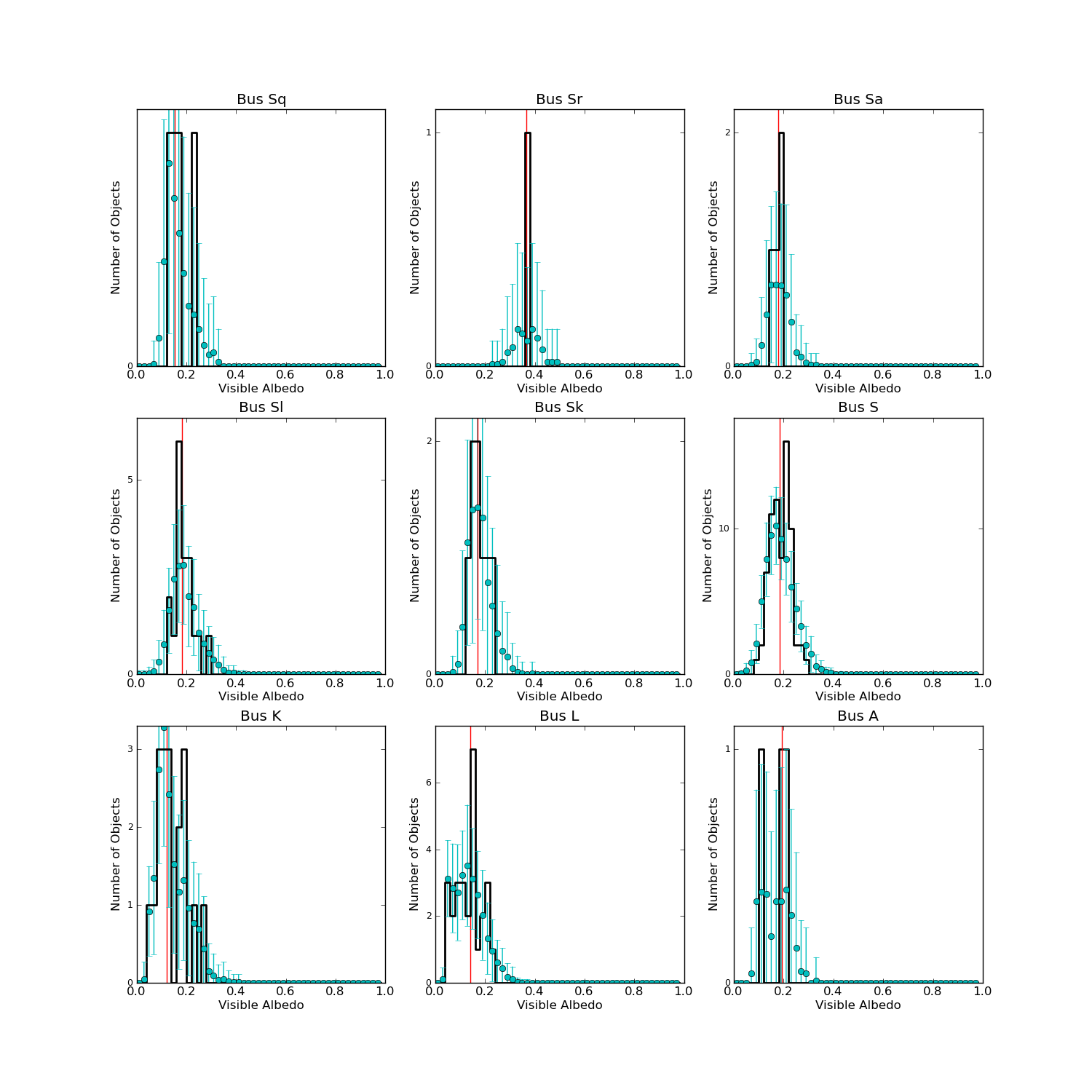}
\caption{\label{fig:BinzelS} NEOWISE-derived $p_{V}$ for S-complex asteroids with diameters larger than 30 km classified using the Bus system are separated into S, Sa, Sk, Sl, Sq and Sr classes; we also show the albedos of objects in the K, L and A classes here.  All S-type asteroids have fairly similar albedo distributions. In \citet{DeMeo09}, the Sa, Sk, and Sl classes have been superceded and are no longer used.}
\end{figure}

\begin{figure}
\figurenum{11}
\includegraphics[width=6in]{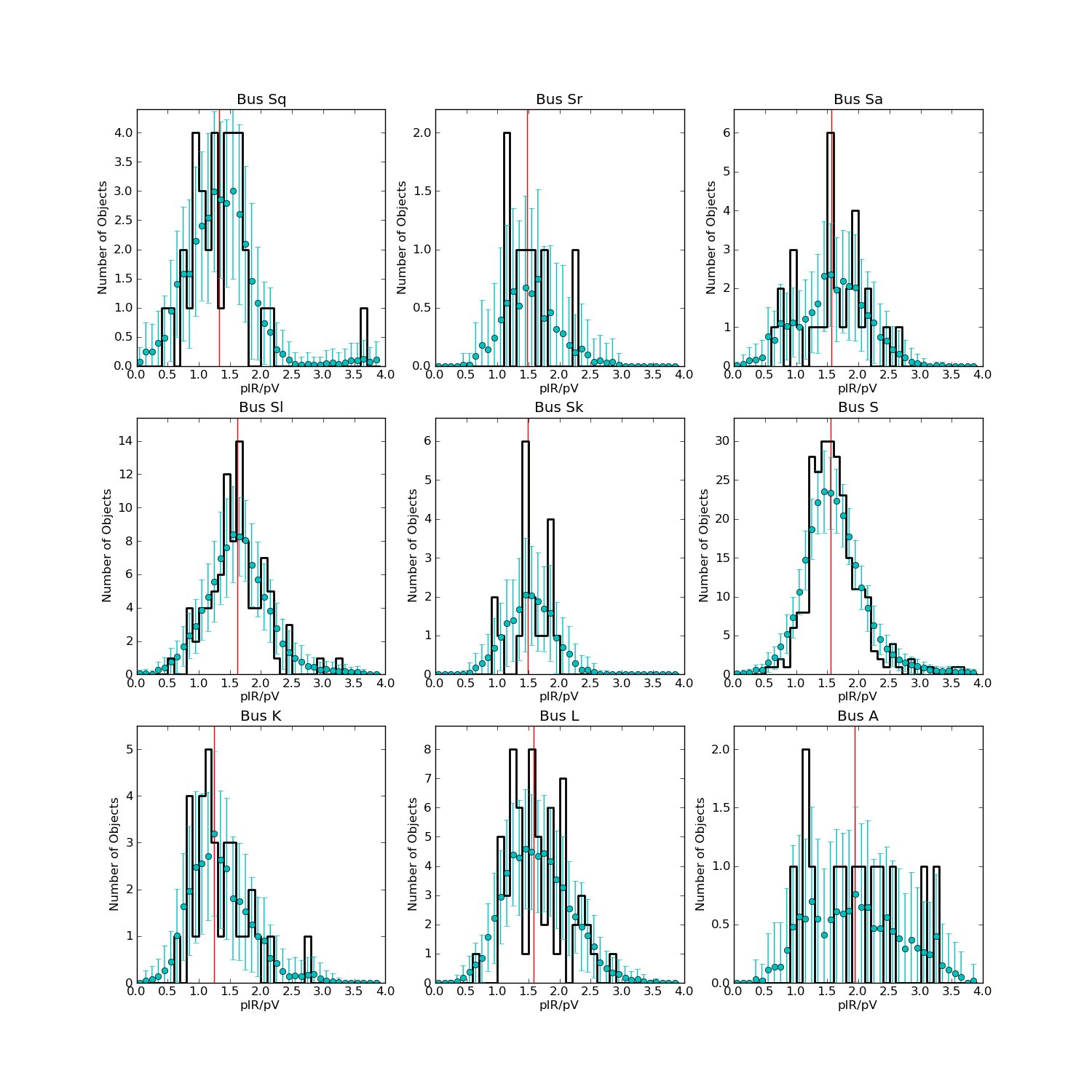}
\caption{\label{fig:BinzelS_pIR} NEOWISE-derived $p_{IR}/p_{V}$ for S-complex asteroids classified using the Bus system. Classes with steeper, redder VNIR slopes tend to have somewhat higher $p_{IR}/p_{V}$ values.}
\end{figure}

\emph{S-Complex.} As expected from \citet{StuartBinzel} and others, the S types observed by NEOWISE tend to have systematically higher albedos than the C types for the Bus, Tholen and DeMeo classification schemes, although they span a fairly wide range. The Bus and Bus-DeMeo taxonomic classification schemes split the S-complex into a number of different sub-classes based on their visible and/or near-infrared slopes and absorption features.  Figures \ref{fig:BinzelS} and \ref{fig:BinzelS_pIR} show the breakdown of  \pv\ and \irfactor, respectively, for the subtypes with diameters larger than 30 km within the Bus S-complex: S, Sa, Sk, Sl, Sr, and Sq, along with the K, L and A types.  The distribution of \pv\ is similar for all of these subtypes; any subtle differences are likely attributable to statistically small numbers of objects for some of the subtypes, with the exception of the K types, which appear to have a somewhat lower albedo as noted in \citet{Tedesco89}.  In the distribution of \irfactor, however, we note some slight differences among subclasses, with the S, Sl and L types showing a slightly higher mean value of \irfactor\ than the Sq, Sk and K types.  According to \citet{Bus}, the S, Sl and L types have redder slopes than the Sk, Sq, and K types.  As with the C, D and T types, redder VNIR slopes correlate with higher $p_{IR}/p_{V}$, possibly indicating that the red slope continues out to 3-4 $\mu$m. However, in general, $p_{V}$ and the $p_{IR}/p_{V}$ ratio of most of the Bus S-complex subtypes are similar.  \citet{DeMeo09} creates a new spectral sequence for the S complex that supercedes the Bus S complex; in the Bus-DeMeo scheme, the Bus Sa disappears, the Bus Sr is converted to the Bus-DeMeo Sa, and the Bus Sl and Sk classes are eliminated.  In the future, all of the $\sim$230 asteroids with these classifications may be redesignated according to the newer Bus-DeMeo system.  Figures \ref{fig:BusDemeopV} and \ref{fig:BusDemeoIR} show the albedo and \irfactor\ distributions for objects with diameters larger than 30 km and more than a handful of objects per taxonomic class.  

It has been asserted that Q type asteroids are the un-space-weathered cousins of the S-type asteroids, with the Bus-DeMeo Sq subtype representing an intermediate state between S and Q types \citep{DeMeo09}.  In the Bus-DeMeo system, types with a w (e.g. Sw, Sqw, Srw) are versions of types with steeper and redder VNIR slopes; \citet{DeMeo09} attribute this reddening to the effects of space weathering.  Space weathering is thought to darken and redden surfaces of airless bodies exposed to radiation; \citet{Chapman04} and \citet{Clark02} give overviews of the subject.  We have observed 65 Main Belt S types classified according to the Bus-DeMeo system and 26 Main Belt asteroids classified as Sw. The S types have a median \pv $=$0.224$\pm$0.013 with a standard deviation of 0.068, while the Sw types have a median \pv $=$0.239$\pm$0.012 with a standard deviation of 0.095 (see Figure \ref{fig:demeo_diam_alb}).  This result suggests that if space weathering is at work on the Sw types, it does not make their surfaces darken; it is also possible that these objects are not actually weathered, or that compositional or surface morphology variations such as differences in regolith particle size creates problems in the comparison between these two groups.  We observed two NEOs classified as Q type, (2102) and (5143), and these objects' albedos are 0.214$\pm$0.095 and 0.227$\pm$0.054 respectively.  With a sample of only two objects, it is difficult to make a statistically meaningful comparison to the S types, although the albedos are entirely consistent with them. We have only three and six Bus-DeMeo Sq and Sqw types, respectively, but their albedos are similar to the S types (see Table 1).  If the Sw and Q types that we observed are space weathered, the process is not affecting their albedos in the predicted manner. Furthermore, in \citet{Masiero}, we found that asteroids in the 5.8 Myr old Karin family have lower albedos than the much older Koronis family, from which the Karin family is thought to originate \citep{Nesvorny}. Determination of asteroid VNIR spectral slopes used by the Bus and Bus-DeMeo systems can be complicated by instrumental effects as described in \citet{Gaffey02} and by reddening of the observed VNIR slopes due to phase effects \citep{Gradie}.  All of these results suggest that the picture of space weathering is complicated, either by compositional variation, variable surface properties, or observational effects.  

\begin{figure}
\figurenum{12}
\includegraphics[width=6in]{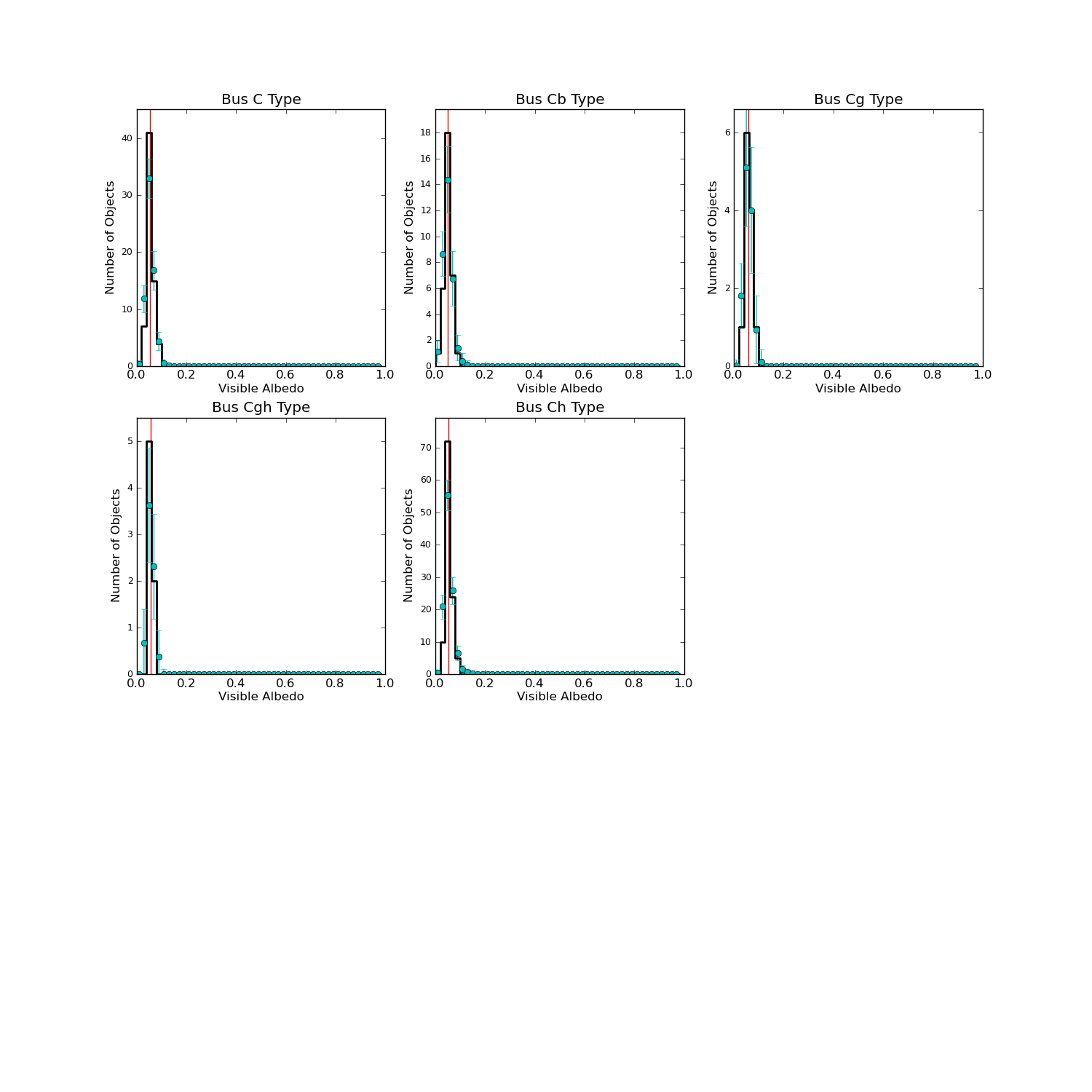}
\caption{\label{fig:BinzelC} NEOWISE-derived $p_{V}$ for C-complex asteroids with diameters larger than 30 km classified using the Bus system are separated into B, C, Cb, Cg, Cgh, and Ch classes.}
\end{figure}

\begin{figure}
\figurenum{13}
\includegraphics[width=6in]{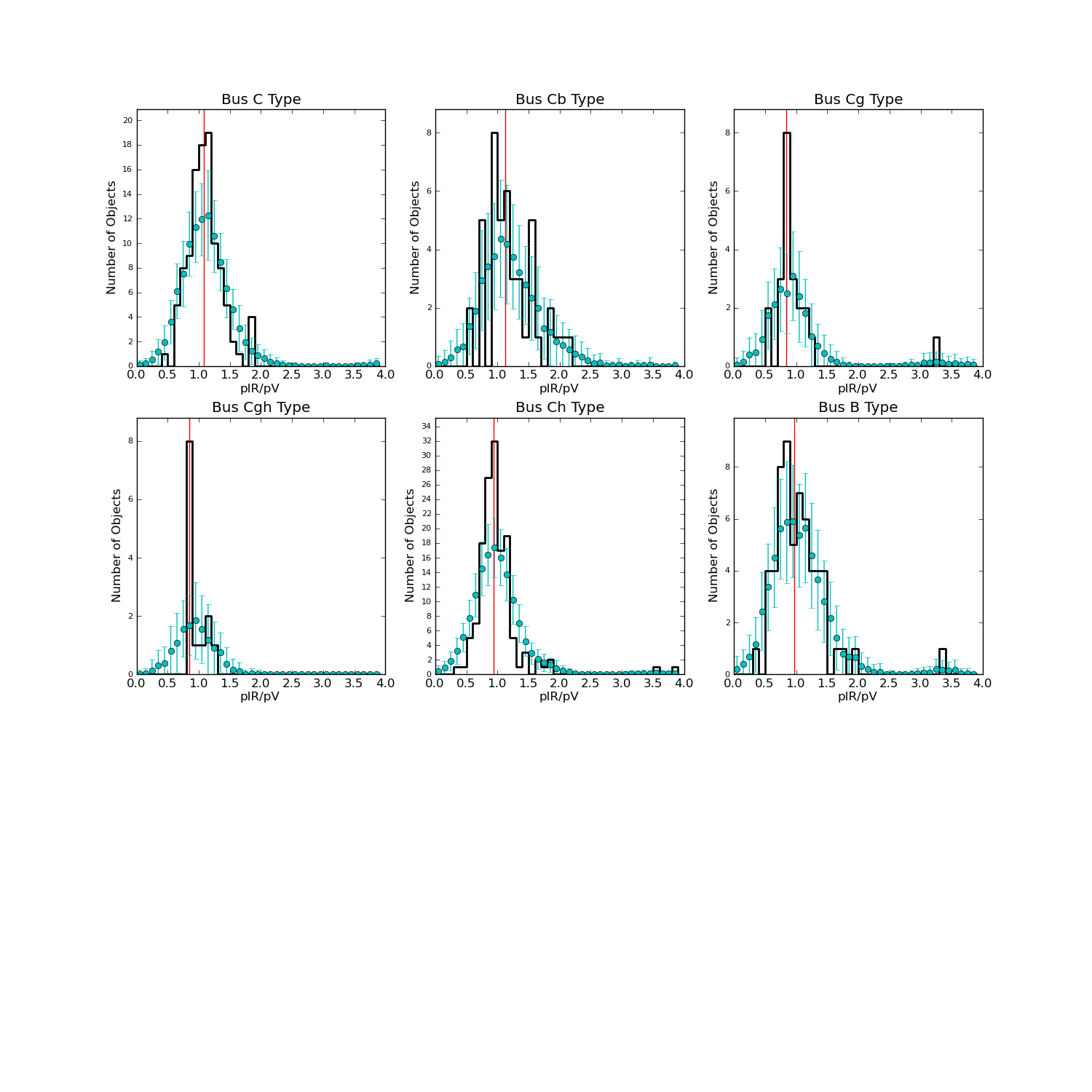}
\caption{\label{fig:BinzelC_pIR} NEOWISE-derived $p_{IR}/p_{V}$ ratio for C-complex asteroids classified using the Bus system are separated into B, C, Cb, Cg, Cgh, and Ch classes.  The B type asteroids show a somewhat lower $p_{IR}/p_{V}$ ratio than the C type asteroids, and this is possibly caused by their somewhat blue VNIR slope extending out to 3-4 $\mu$m.}
\end{figure}

\begin{figure}
\figurenum{14}
\includegraphics[width=6in]{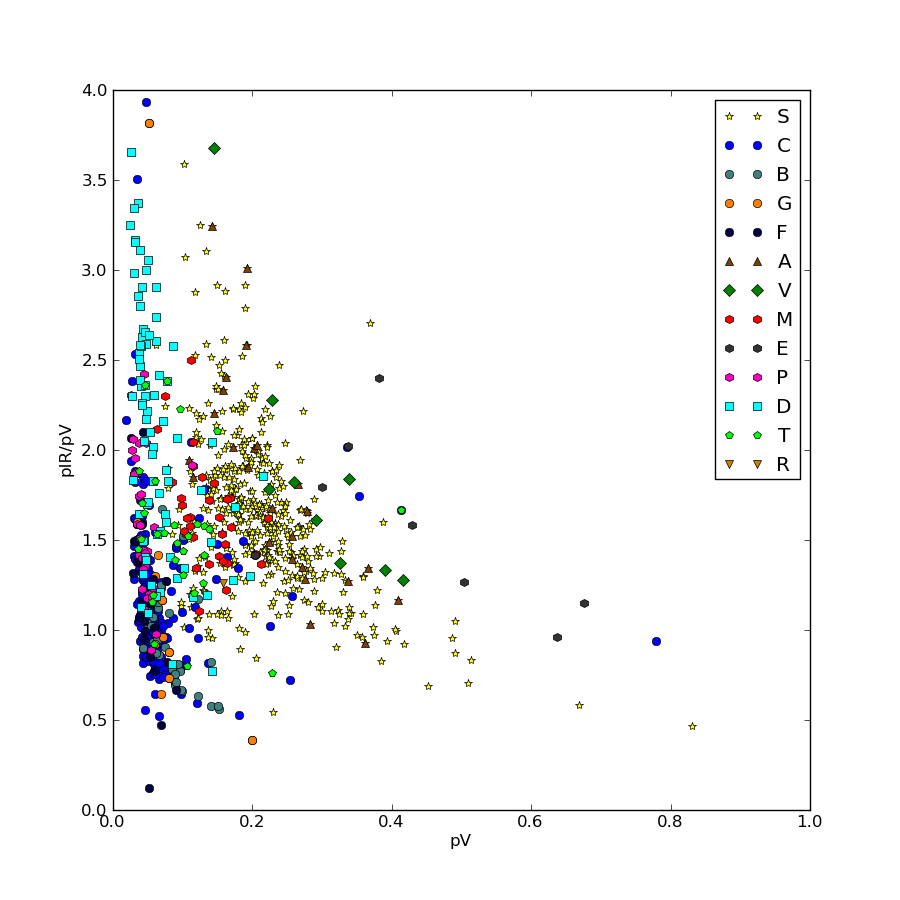}
\caption{\label{fig:scatter_tholen} NEOWISE-derived ratio $p_{IR}/p_{V}$ vs. $p_{V}$ for asteroids observed and classified according to the Tholen taxonomic classification scheme.  Only asteroids for which $p_{IR}/p_{V}$ could be fitted are included in this plot. }
\end{figure}

\begin{figure}
\figurenum{15}
\includegraphics[width=6in]{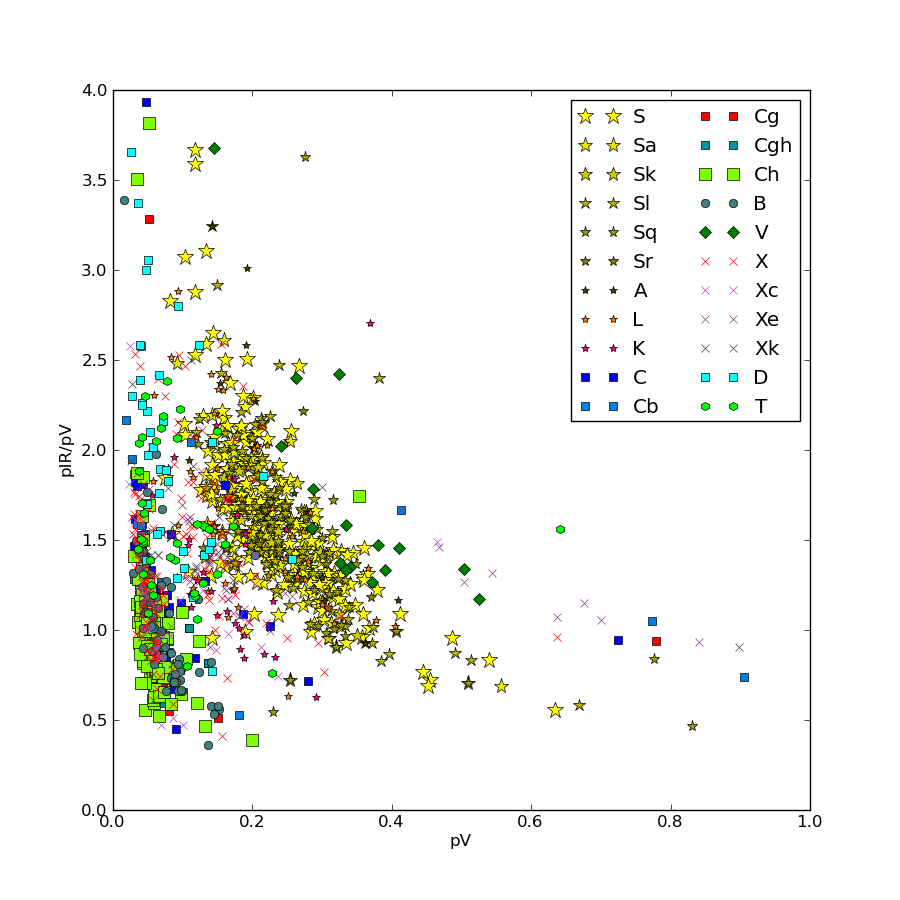}
\caption{\label{fig:scatter_binzel} NEOWISE-derived ratio $p_{IR}/p_{V}$ vs. $p_{V}$ for asteroids observed and classified according to the system of \citet{Bus}.  Only asteroids for which $p_{IR}/p_{V}$ could be fitted are included in this plot.}
\end{figure}

\emph{C-Complex.} The NEOWISE $p_{V}$ and $p_{IR}/p_{V}$ for the Bus and Bus-DeMeo C-complex asteroids are shown for the B, C, Cb, Ch, Cg, and Cgh types in Figures \ref{fig:BinzelC} and \ref{fig:BinzelC_pIR}.  In all three taxonomic schemes, the B, C, D and T types all have similarly low $p_{V}$ values, $\sim$0.05.  In the visible and near-infrared, C type asteroids are characterized by relatively flat spectra between 0.4 and 1.0 $\mu$m with few, if any, absorption features.  In the Bus and Bus-DeMeo taxonomic schemes, the C-complex is differentiated by the presence or absence of a broad absorption feature near 0.7 $\mu$m; \citet{Bus} divided objects with and without this feature into five further subclasses (C, Cb, Cg, Ch, Cgh) depending additionally on the slope of the spectrum shortward of 0.55 $\mu$m.  By contrast, the T and D types have featureless spectra that are nevertheless characterized by moderate and steep red VNIR slopes, respectively, whereas the B types have a slightly blue slope. The quantity \irfactor\ can be extremely useful for differentiating asteroids. While the B, C, D and T types all have extremely similar \pv, their \irfactor\ ratios are significantly different.  As shown in Table 1, the T and D types have increasingly larger values of $p_{IR}/p_{V}$, indicating that the steep slopes observed between visible and near-infrared wavelengths are likely to continue through the 3.4 and 4.6 $\mu$m WISE bands.  A discussion of the possible materials responsible for the spectral appearance of the primitive Trojan asteroids out to 4 $\mu$m can be found in \citet{Emery}.   Figures \ref{fig:scatter_tholen} and \ref{fig:scatter_binzel} illustrate the utility that $p_{IR}/p_{V}$ can provide for distinguishing various taxonomic types (including the many subclasses within each complex) from one another in both the Tholen and Bus schemes.

\begin{figure}
\figurenum{16}
\includegraphics[width=6in]{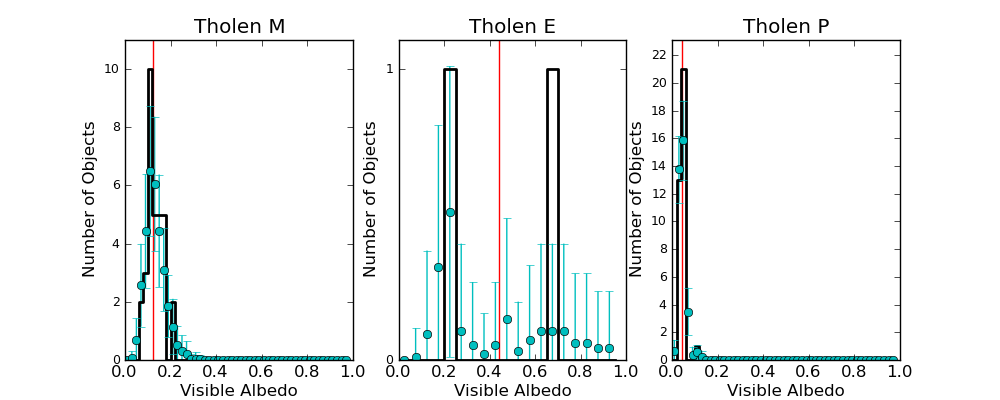}
\caption{\label{fig:TholenX}The E, M and P classes that make up the Tholen X type are distinguishable by albedo, as expected from Tholen's definition. }
\end{figure}

\begin{figure}
\figurenum{17}
\includegraphics[width=6in]{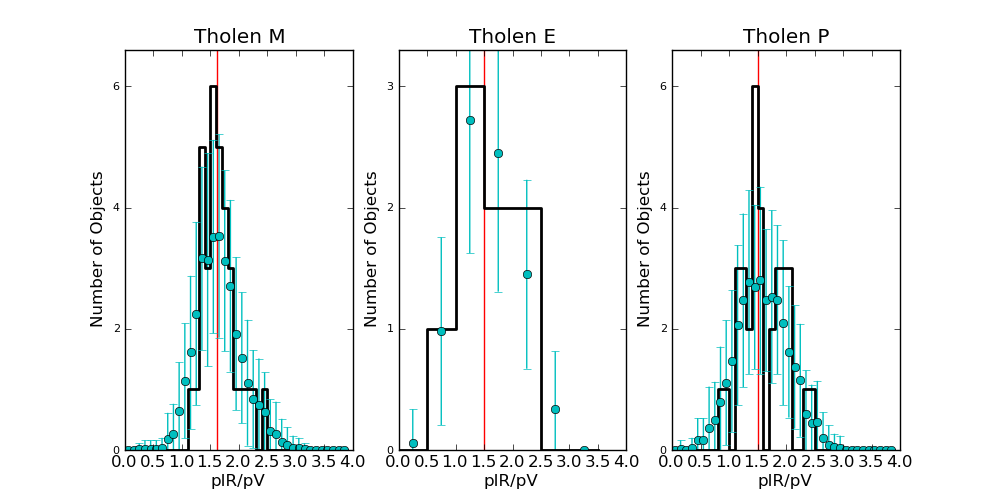}
\caption{\label{fig:TholenX_pIR} The E, M and P classes that make up the Tholen X type are not distinguishable by $p_{IR}/p_{V}$. }
\end{figure}

\begin{figure}
\figurenum{18}
\includegraphics[width=6in]{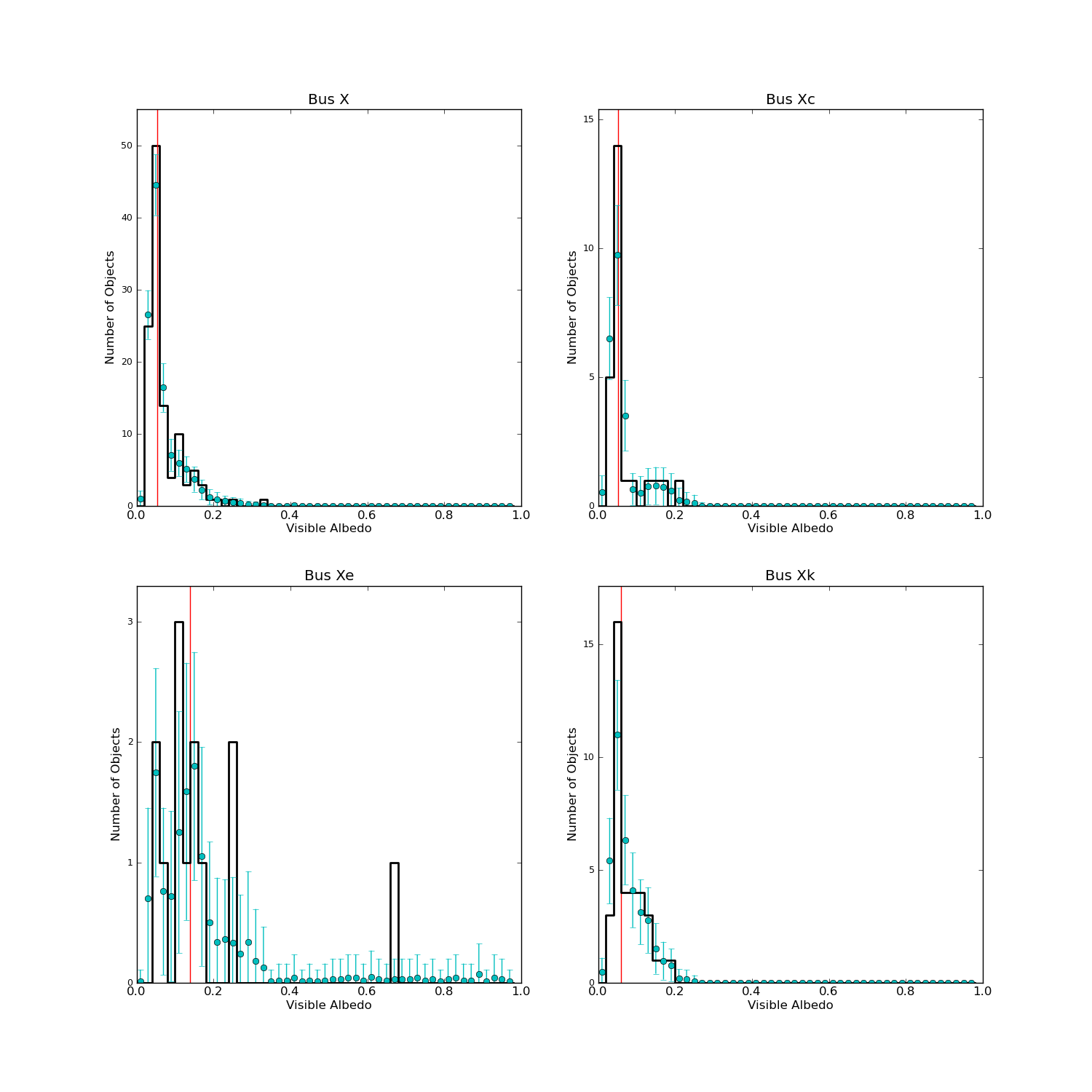}
\caption{\label{fig:BinzelX} X-complex asteroids classified using the Bus system are separated into X, Xc, Xe and Xk classes; unlike the Tholen X classification, the Bus and Bus-DeMeo schemes do not use albedo. This ambiguity with respect to albedo is reflected in the similarity in the average albedos for the X, Xc, Xe and Xk classes, although Xe is somewhat higher (see Table 1). }
\end{figure}

\begin{figure}
\figurenum{19}
\includegraphics[width=6in]{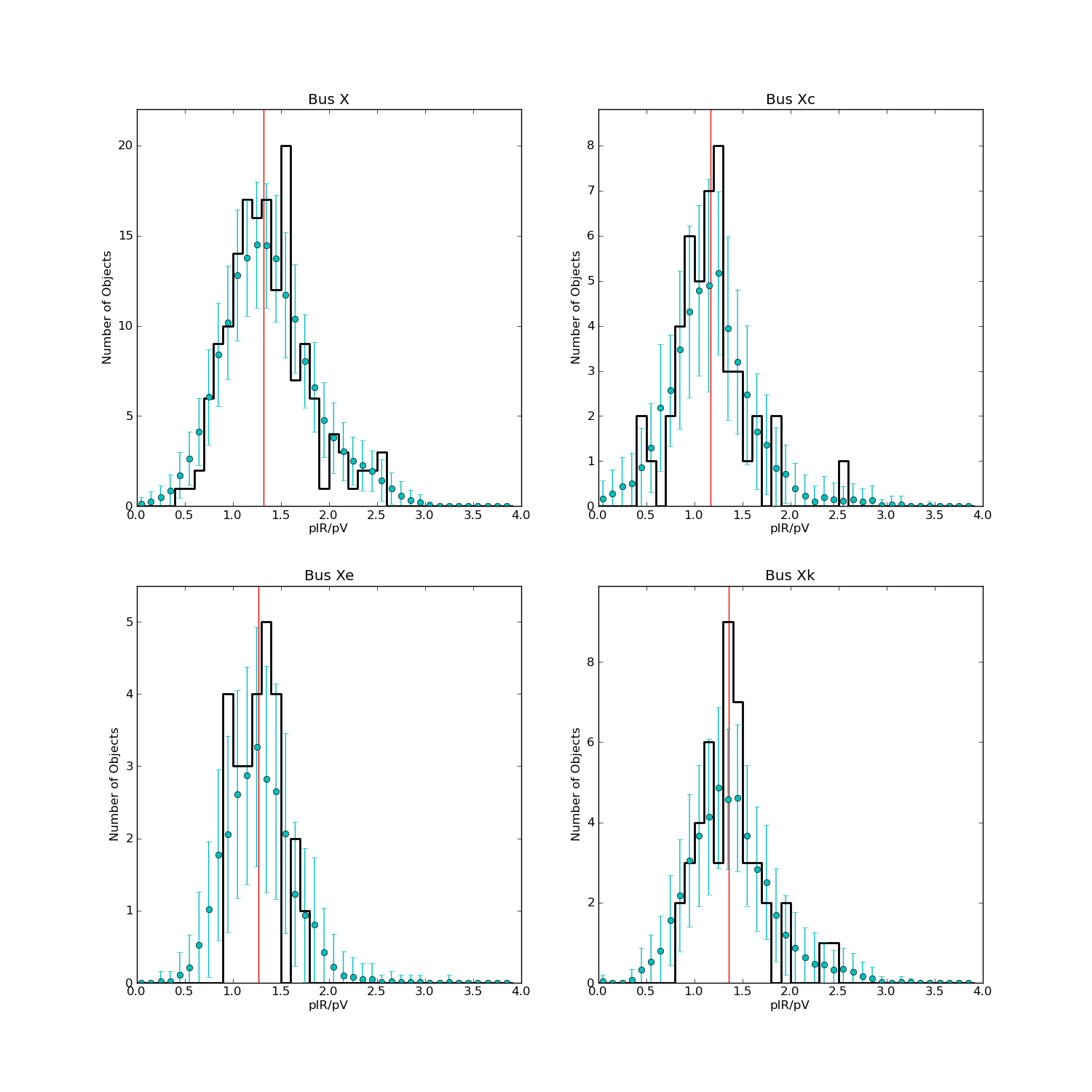}
\caption{\label{fig:BinzelX_pIR} The X, Xc, Xe, and Xk Bus classes within the X-complex have similar values of $p_{IR}/p_{V}$.   }
\end{figure}

\emph{X-Complex.}  The Tholen, Bus and Bus-DeMeo X types span a wide range of albedos, from $\sim$0.07-$>$0.6.  This wide range is to be expected, as the Tholen X type (from which the Bus and Bus-DeMeo X types are derived) is comprised of E, M and P asteroids which are distinguished on the basis of their albedos (Figure \ref{fig:TholenX}; Figure \ref{fig:TholenX_pIR} shows the ratio $p_{IR}/p_{V}$ for the Tholen X types). The albedo distribution of the asteroids with Tholen X classifications and Bus X types follow a distribution that reflects the distribution observed in the Main Belt \citep{Masiero}.  Since neither the Bus nor the Bus-DeMeo taxonomic systems use albedo for classification, it is perhaps unsurprising that when their X-complex objects are broken down into the X, Xc, Xe, and Xk subclasses (Figure \ref{fig:BinzelX} and \ref{fig:BinzelX_pIR}), $p_{V}$ and $p_{IR}/p_{V}$ appear to be similar for all of them.  However, both Bus and Bus-DeMeo recognize the Xe class as being indicative of the high albedo E types in the Tholen taxonomy. In Table 2, we assign Tholen-style E, M and P classifications to X-complex objects that do not already have E, M or P classification based on their NEOWISE preliminary albedos.

\emph{Others.} The V type asteroid class was first proposed by \citet{Tholen84}; since then, a number of Vestoids have been identified both dynamically and spectroscopically as being related to the parent body (4) Vesta.   As expected, V-type asteroids have higher albedos, on average, than the S-complex asteroids.  The few asteroids classified as O-types by \citet{Bus} fall within the broad range of the S-complex.  

As noted above, we have assumed that $p_{W1}=p_{W2}$; future work will attempt to determine whether or not the albedo at 3.4 and 4.6 $\mu$m really is the same.  To test the degree to which this assumption affects the resulting diameters, \pv\ and \irfactor\ values, we recomputed the thermal fits without using band $W2$.  This analysis resulted in no significant changes to either diameter, \pv, or \irfactor; almost all fits agreed to within $\pm$10\% of their original values.  This result is perhaps not surprising.  The diameter is most strongly influenced by the thermal emission-dominated bands $W3$ and $W4$ for Main Belt asteroids, which make up the vast majority of our sample. Visible albedo and \irfactor\ are more heavily influenced by band $W1$ than $W2$, since this band consists almost entirely of reflected sunlight, while band $W2$ most always has less reflected light than thermal emission. 

There are a number of different possible causes of the variations we observe in \irfactor\ for different objects.  Even a cursory examination of mineralogical and meteorite databases yields a wealth of different materials with features in wavelengths covered by bands $W1$ and $W2$.  \citet{Gaffey02} and references therein summarize some of the possible causes of features in these wavelength regimes: a 3 $\mu$m feature indicating the presence of hydration caused by the fundamental O-H stretch bands of H$_{2}$O; anhydrous assemblages of mafic silicates containing structural OH; possible fluid inclusions; or the presence of troilite.  \citet{Rivkin2000} carried out spectrophotometric observations of asteroids in the 1.2 to 3.5 $\mu$m region and found evidence of absorption at 3 $\mu$m; they conclude that these are produced by hydrated minerals. Of the 27 M-type asteroids studied in \citet{Rivkin2000}, 10 showed evidence of an absorption feature at 3 $\mu$m.  With NEOWISE, we observed seven of these: (22) Kalliope, (77) Frigga, (110) Lydia, (129) Antigone, (135) Hertha, (136) Austria, and (201) Penelope. As \citet{Rivkin} reports that the depth of the absorption band at 3 $\mu$m is only $\sim$10-20\% of the continuum flux over a fairly narrow range of wavelengths, we conclude that it would be unlikely to show a detectable change to $p_{IR}/p_{V}$ given that the $W1$ bandpass extends from $2.8 - 3.8 \mu$m \citep{Wright}. These seven objects have a median $p_{V}=0.157\pm0.010$, and their median $p_{IR}/p_{V}=1.572\pm0.050$.  This latter matches the $p_{IR}/p_{V}$ found for the 33 M type asteroids shown in Figure \ref{fig:TholenX_pIR}, which have a median \irfactor\ of 1.623$\pm$0.051 and standard deviation of 0.291.   \citet{Merenyi} show a number of additional asteroids with evidence of absorption at 3 $\mu$m, including the C type asteroid (1467) Mashona, which is given as having a band depth of 88\%. We find that this asteroid has \irfactor$\sim$0.9; however, this value is entirely in line with the rest of the C type asteroids. It is possible, even likely, that the spread in \irfactor\ that we observe could represent nothing more than the natural variation in spectral slope within the various spectral classes.

As discussed above and demonstrated by Figures \ref{fig:tholen_diam_alb} and \ref{fig:binzel_diam_alb}, caution must be exercised when attempting to generalize the fractional population results presented herein to all NEOs or Main Belt asteroids.  The objects selected for taxonomic classification were chosen on the basis of their discovery by visible light surveys, so the selection is inherently biased in favor of high albedo objects. Although \citet{StuartBinzel} compute the relative fractions of asteroids of various taxonomic types observed throughout the solar system, we do not attempt such an undertaking here.  \citet{Thomas11} compare the albedo distributions of NEOs found using 3.6 and 4.5 $\mu$m imaging from the \emph{Spitzer Space Telescope} to the albedo distributions of Main Belt asteroids; while they find that the NEO albedos are higher than Main Belt albedos for various spectral types this result is perhaps not surprising given that the Warm Spitzer sample was drawn from optically selected NEOs.  We have observed relatively few NEOs with taxonomic classifications with WISE and will have to wait until more taxonomic classifications are in hand before making comparisons between NEOs and Main Belt asteroids. The point of such an exercise would be to determine the relative numbers, compositions, sizes and distribution of asteroids of various populations throughout the solar system. We have computed the debiased size and albedo distributions of the NEOs in \citet{Mainzer11d}, and we are computing similar distributions for the Main Belt asteroids, Trojans and comets.  By working with the entire NEOWISE dataset, these works can provide a more direct accounting for the distribution of asteroid albedos and sizes for different populations.

\section{Conclusions}
With the advent of a large, thermal infrared survey of asteroids throughout the Solar System, the NEOWISE dataset offers the opportunity to study the relationship between albedo and various spectral features with unprecedented clarity.  We have computed the preliminary observed range of possible albedos for the various classes using $\sim$1800 NEOs and Main Belt asteroids we observed with NEOWISE. This may allow important physical parameters to be used in the refinement of existing taxonomic classification schemes, or perhaps to allow objects of different types to be more readily distinguished from one another.  Although reasonably good separation between the two main S and C taxonomic complexes can be observed for diameters $>$30 km, where the visible light surveys that found them are largely complete, all taxonomic types and subtypes show an uptick in average albedos at smaller sizes. We attribute this uptick to strong selection biases against finding and classifying small, dark objects with VNIR spectroscopy.  For objects $>$30 km, it is clear that a median albedo can be used based on taxonomic classification.  One could assume that the median albedos for smaller sizes are similar, but the strong selection biases against small, low albedo objects in this study preclude us from deriving or verifying that these median albedos extend to smaller sizes.  Due to the same selection biases, we are thus unable to comment on the relationship between size, albedo and space weathering, although comparison between S and Sw Bus-DeMeo types shows no evidence that the Sw types are darker at any observed size scales.  The two Q type objects we observed have nearly identical albedos to the S types, but a larger number of classified Q types from our dataset is needed to confirm this result.  We do not observe any major distinctions in albedo among the S subtypes and C subtypes in the Bus and Bus-DeMeo systems.  From an albedo perspective, Figures \ref{fig:tholen_diam_alb}, \ref{fig:binzel_diam_alb}, and \ref{fig:demeo_diam_alb} make the Tholen system stand out as the cleanest. While the Tholen system uses albedo to separate the X types into E, M, and P classes, albedo is not used to define the remainder of the classes in the Tholen system.

There is a strong selection bias in the taxonomic classification schemes and average albedos presented here (clearly in Figures \ref{fig:tholen_diam_alb}, \ref{fig:binzel_diam_alb}, \ref{fig:demeo_diam_alb}) and by other observers.  First, since all the objects selected for taxonomic classification have been drawn from visible light surveys, the relative fractional abundance of objects with particular taxonomic types is biased toward higher fractions of high albedo objects.  Second, within a particular taxonomic class, lower albedo objects are less likely to have been observed because they tend to be fainter in visible light: this will skew the average albedo for a particular taxonomic type higher.  Because of these biases, when the average albedo is used to convert from absolute $H$ magnitude to size, artificially smaller sizes for asteroids will be found.  This speaks to the need to assemble a sample of objects with taxonomic classifications that are drawn from the NEOWISE thermal infrared survey to mitigate biases against low albedo objects.   

With the four infrared wavelengths given by the WISE dataset, we are able to derive the ratio of the albedo at 3.4 and 4.6 $\mu$m to the visible albedo.  We have shown that taxonomic types with steeply red spectral slopes in VNIR wavelengths tend to have higher $p_{IR}/p_{V}$ values.  We hypothesize that this is caused by the fact that the spectral slopes continue to rise from visible through the near-infrared to the $W1$ and $W2$ wavelengths for these objects.  For example, we have shown that spectral types T and D can be distinguished from the C types by examining their $p_{IR}/p_{V}$, even though they have virtually identical $p_{V}$.  Subclasses within the S and C complexes generally have similar visible albedos and largely similar $p_{IR}/p_{V}$ ratios.  However,  $p_{IR}/p_{V}$ can only be computed when a sufficiently high fraction of reflected sunlight is present in either bands $W1$ or $W2$.  The bias against low albedo objects is present in the determination of $p_{IR}$, in that dark objects are less likely to have enough reflected sunlight in bands $W1$ or $W2$ to allow $p_{IR}$ to be computed.  As before, we caution against generalizing the average $p_{IR}/p_{V}$ values we have given here to entire populations or classes of objects in light of the presence of these biases.

This work shows that the NEOWISE dataset offers a new means of exploring the connections between taxonomic classifications derived from VNIR spectroscopy and spectrophotometry.  Future work will explore the relationship between visible albedo and the 3-4 $\mu$m albedo to VNIR spectroscopic properties in greater detail.  The value of the NEOWISE dataset will only be enhanced by the acquisition of additional visible and near-infrared ancillary data. More data would be beneficial for two reasons.  First, we require a measurement of $H$ in order to determine $p_{V}$ and $p_{IR}/p_{V}$, so more accurate $H$ and $G$ values will result in more accurate albedos.  Second, by obtaining taxonomic classification of low albedo objects drawn from the NEOWISE sample, we can reduce the bias within each taxonomic class against lower albedo objects. With the NEOWISE dataset, we now have access to a means of directly computing debiased size and albedo distributions that are not as subject to the biases against low albedo objects as objects selected for classification and study by visible light surveys.  

\section{Acknowledgments}

\acknowledgments{This publication makes use of data products from the \emph{Wide-field Infrared Survey Explorer}, which is a joint project of the University of California, Los Angeles, and the Jet Propulsion Laboratory/California Institute of Technology, funded by the National Aeronautics and Space Administration.  This publication also makes use of data products from NEOWISE, which is a project of the Jet Propulsion Laboratory/California Institute of Technology, funded by the Planetary Science Division of the National Aeronautics and Space Administration.  We gratefully acknowledge the extraordinary services specific to NEOWISE contributed by the International Astronomical Union's Minor Planet Center, operated by the Harvard-Smithsonian Center for Astrophysics, and the Central Bureau for Astronomical Telegrams, operated by Harvard University.  We thank the paper's referee, Prof. Richard Binzel, for his helpful contributions.  We also thank the worldwide community of dedicated amateur and professional astronomers devoted to minor planet follow-up observations. This research has made use of the NASA/IPAC Infrared Science Archive, which is operated by the Jet Propulsion Laboratory, California Institute of Technology, under contract with the National Aeronautics and Space Administration.}

\clearpage
 
\begin{deluxetable}{lllllllllllll}
\tabletypesize{\tiny}
\tablecolumns{13}
\tablecaption{Median values of $p_{V}$ and $p_{IR}/p_{V}$ for various taxonomic types using NEOWISE cryogenic observations of NEOs and Main Belt asteroids. The medians, standard deviations of the mean (SDOM) and standard deviations (SD) given were computed simply by taking the median and standard deviation of all the objects with a particular classification; however, a more complete picture of the distribution and full range of albedos within a taxonomic class is given in the figures, which show the shapes of the distributions.  Note that while $p_{V}$ was fitted for all objects in the table, if an asteroid did not have a sufficient number of observations in $W1$ or $W2$,  $p_{IR}/p_{V}$ could not be fit.  Therefore, not all taxonomic types have the same number of objects with $p_{V}$ and $p_{IR}/p_{V}$. Only objects with fitted $p_{IR}/p_{V}$ were used in the computation of median $p_{IR}/p_{V}$ given here.}
\tablehead{
\colhead{Class} & \colhead{N ($p_{V}$)} & \colhead{Med. $p_{V}$} & \colhead{SD} & \colhead{SDOM} & \colhead{Min} & \colhead{Max} & \colhead{N ($p_{IR}/p_{V}$)} & \colhead{Med. $p_{IR}/p_{V}$} & \colhead{SD} & 
\colhead{SDOM} & \colhead{Min} & \colhead{Max}
}
\startdata
Bus A & 14 & 0.234 & 0.084 & 0.022 & 0.110 & 0.410 & 13 & 1.943 & 0.697 & 0.193 & 0.926 & 3.244\\
Bus B & 79 & 0.075 & 0.087 & 0.010 & 0.016 & 0.720 & 60 & 0.970 & 0.441 & 0.057 & 0.363 & 3.387\\
Bus C Complex & 367 & 0.058 & 0.086 & 0.004 & 0.018 & 0.905 & 312 & 0.994 & 0.411 & 0.023 & 0.390 & 3.934\\
Bus C & 128 & 0.059 & 0.073 & 0.006 & 0.031 & 0.725 & 107 & 1.088 & 0.379 & 0.037 & 0.448 & 3.934\\
Bus Cb & 53 & 0.055 & 0.154 & 0.021 & 0.018 & 0.905 & 44 & 1.124 & 0.385 & 0.058 & 0.528 & 2.167\\
Bus Cg & 27 & 0.067 & 0.134 & 0.026 & 0.037 & 0.769 & 22 & 0.844 & 0.531 & 0.113 & 0.511 & 3.281\\
Bus Cgh & 15 & 0.065 & 0.032 & 0.008 & 0.044 & 0.137 & 13 & 0.848 & 0.149 & 0.041 & 0.804 & 1.286\\
Bus Ch & 163 & 0.056 & 0.036 & 0.003 & 0.031 & 0.353 & 143 & 0.939 & 0.398 & 0.033 & 0.390 & 3.814\\
Bus D & 44 & 0.075 & 0.055 & 0.008 & 0.026 & 0.257 & 37 & 1.974 & 0.631 & 0.104 & 0.773 & 3.653\\
Bus K & 34 & 0.157 & 0.067 & 0.011 & 0.054 & 0.370 & 32 & 1.248 & 0.432 & 0.076 & 0.628 & 2.704\\
Bus L & 72 & 0.176 & 0.082 & 0.010 & 0.030 & 0.405 & 63 & 1.583 & 0.600 & 0.076 & 0.631 & 4.829\\
Bus O & 3 & 0.227 & 0.067 & 0.039 & 0.178 & 0.339 & 1 & 2.084 & 0.000 & 0.000 & 2.084 & 2.084\\
Bus Q & 1 & 0.214 & 0.000 & 0.000 & 0.214 & 0.214 & 0 & 0.000 & 0.000 &   nan & 0.000 & 0.000\\
Bus R & 0 &   nan & 0.000 &   nan & 0.000 & 0.000 & 2 & 1.309 & 0.046 & 0.032 & 1.264 & 1.355\\
Bus S Complex & 531 & 0.234 & 0.088 & 0.004 & 0.085 & 0.830 & 433 & 1.554 & 0.446 & 0.021 & 0.467 & 3.664\\
Bus S & 312 & 0.227 & 0.078 & 0.004 & 0.085 & 0.635 & 256 & 1.557 & 0.432 & 0.027 & 0.557 & 3.664\\
Bus Sa & 39 & 0.230 & 0.099 & 0.016 & 0.092 & 0.557 & 30 & 1.563 & 0.498 & 0.091 & 0.689 & 2.613\\
Bus Sk & 22 & 0.215 & 0.059 & 0.013 & 0.133 & 0.365 & 19 & 1.490 & 0.292 & 0.067 & 0.956 & 1.907\\
Bus Sl & 102 & 0.230 & 0.087 & 0.009 & 0.120 & 0.669 & 94 & 1.616 & 0.442 & 0.046 & 0.586 & 3.244\\
Bus Sq & 54 & 0.282 & 0.127 & 0.017 & 0.097 & 0.830 & 36 & 1.329 & 0.546 & 0.091 & 0.467 & 3.627\\
Bus Sr & 14 & 0.282 & 0.072 & 0.019 & 0.210 & 0.438 & 7 & 1.478 & 0.350 & 0.132 & 1.122 & 2.217\\
Bus T & 42 & 0.086 & 0.095 & 0.015 & 0.036 & 0.641 & 38 & 1.500 & 0.407 & 0.066 & 0.762 & 2.384\\
Bus V & 24 & 0.350 & 0.109 & 0.022 & 0.146 & 0.653 & 16 & 1.463 & 0.625 & 0.156 & 1.170 & 3.676\\
Bus X,Xc,Xe,Xk & 313 & 0.074 & 0.153 & 0.009 & 0.024 & 0.896 & 279 & 1.297 & 0.394 & 0.024 & 0.413 & 2.587\\
Bus X & 178 & 0.062 & 0.115 & 0.009 & 0.028 & 0.896 & 163 & 1.323 & 0.419 & 0.033 & 0.413 & 2.587\\
Bus Xc & 54 & 0.086 & 0.162 & 0.022 & 0.024 & 0.848 & 47 & 1.170 & 0.366 & 0.053 & 0.472 & 2.578\\
Bus Xe & 31 & 0.174 & 0.238 & 0.043 & 0.043 & 0.841 & 26 & 1.270 & 0.221 & 0.043 & 0.906 & 1.781\\
Bus Xk & 53 & 0.079 & 0.119 & 0.016 & 0.027 & 0.862 & 46 & 1.361 & 0.347 & 0.051 & 0.801 & 2.498\\
Bus-DeMeo A & 5 & 0.191 & 0.034 & 0.015 & 0.110 & 0.207 & 5 & 2.030 & 0.416 & 0.186 & 1.943 & 3.010\\
Bus-DeMeo B & 2 & 0.120 & 0.022 & 0.015 & 0.098 & 0.142 & 1 & 0.575 & 0.000 & 0.000 & 0.575 & 0.575\\
Bus-DeMeo C Complex & 32 & 0.058 & 0.028 & 0.005 & 0.036 & 0.204 & 32 & 1.014 & 0.535 & 0.095 & 0.548 & 3.814\\
Bus-DeMeo C & 9 & 0.050 & 0.006 & 0.002 & 0.047 & 0.063 & 9 & 1.180 & 0.122 & 0.041 & 0.926 & 1.404\\
Bus-DeMeo Cb & 1 & 0.043 & 0.000 & 0.000 & 0.043 & 0.043 & 1 & 1.528 & 0.000 & 0.000 & 1.528 & 1.528\\
Bus-DeMeo Cg & 1 & 0.063 & 0.000 & 0.000 & 0.063 & 0.063 & 1 & 0.950 & 0.000 & 0.000 & 0.950 & 0.950\\
Bus-DeMeo Cgh & 8 & 0.065 & 0.048 & 0.017 & 0.051 & 0.204 & 8 & 0.929 & 0.250 & 0.088 & 0.548 & 1.416\\
Bus-DeMeo Ch & 13 & 0.058 & 0.009 & 0.003 & 0.036 & 0.073 & 13 & 0.961 & 0.790 & 0.219 & 0.557 & 3.814\\
Bus-DeMeo D & 13 & 0.048 & 0.025 & 0.007 & 0.029 & 0.116 & 11 & 2.392 & 0.533 & 0.161 & 1.484 & 3.375\\
Bus-DeMeo K & 11 & 0.130 & 0.058 & 0.018 & 0.080 & 0.291 & 11 & 1.278 & 0.326 & 0.098 & 0.628 & 1.899\\
Bus-DeMeo L & 19 & 0.149 & 0.066 & 0.015 & 0.054 & 0.304 & 16 & 1.220 & 0.315 & 0.079 & 0.631 & 1.885\\
Bus-DeMeo O & 1 & 0.339 & 0.000 & 0.000 & 0.339 & 0.339 & 0 & 0.000 & 0.000 &   nan & 0.000 & 0.000\\
Bus-DeMeo Q & 1 & 0.227 & 0.000 & 0.000 & 0.227 & 0.227 & 0 & 0.000 & 0.000 &   nan & 0.000 & 0.000\\
Bus-DeMeo R & 1 & 0.148 & 0.000 & 0.000 & 0.148 & 0.148 & 1 & 1.264 & 0.000 & 0.000 & 1.264 & 1.264\\
Bus-DeMeo S Complex & 121 & 0.223 & 0.073 & 0.007 & 0.114 & 0.557 & 105 & 1.666 & 0.469 & 0.046 & 0.689 & 3.627\\
Bus-DeMeo S & 66 & 0.211 & 0.068 & 0.008 & 0.114 & 0.456 & 59 & 1.602 & 0.312 & 0.041 & 0.724 & 2.288\\
Bus-DeMeo Sa & 1 & 0.367 & 0.000 & 0.000 & 0.367 & 0.367 & 1 & 1.183 & 0.000 & 0.000 & 1.183 & 1.183\\
Bus-DeMeo Sq & 6 & 0.243 & 0.039 & 0.016 & 0.160 & 0.276 & 6 & 1.867 & 0.695 & 0.284 & 1.573 & 3.627\\
Bus-DeMeo Sqw & 7 & 0.231 & 0.043 & 0.016 & 0.195 & 0.311 & 7 & 1.763 & 0.365 & 0.138 & 0.956 & 2.064\\
Bus-DeMeo Sr & 10 & 0.266 & 0.055 & 0.018 & 0.163 & 0.352 & 7 & 1.541 & 0.383 & 0.145 & 1.165 & 2.424\\
Bus-DeMeo Srw & 2 & 0.279 & 0.051 & 0.036 & 0.227 & 0.330 & 0 & 0.000 & 0.000 &   nan & 0.000 & 0.000\\
Bus-DeMeo Sv & 1 & 0.309 & 0.000 & 0.000 & 0.309 & 0.309 & 0 & 0.000 & 0.000 &   nan & 0.000 & 0.000\\
Bus-DeMeo Svw & 0 &   nan & 0.000 &   nan & 0.000 & 0.000 & 0 & 0.000 & 0.000 &   nan & 0.000 & 0.000\\
Bus-DeMeo Sw & 28 & 0.221 & 0.094 & 0.018 & 0.119 & 0.557 & 25 & 1.790 & 0.632 & 0.126 & 0.689 & 3.244\\
Bus-DeMeo T & 2 & 0.042 & 0.004 & 0.003 & 0.037 & 0.046 & 2 & 1.843 & 0.195 & 0.138 & 1.648 & 2.038\\
Bus-DeMeo V & 8 & 0.362 & 0.100 & 0.035 & 0.242 & 0.526 & 7 & 1.335 & 0.553 & 0.209 & 0.558 & 2.400\\
Bus-DeMeo Vw & 0 &   nan & 0.000 &   nan & 0.000 & 0.000 & 0 & 0.000 & 0.000 &   nan & 0.000 & 0.000\\
\tablebreak
Bus-DeMeo X Complex & 17 & 0.111 & 0.143 & 0.035 & 0.036 & 0.676 & 17 & 1.440 & 0.334 & 0.081 & 1.054 & 2.498\\
Bus-DeMeo X & 3 & 0.047 & 0.060 & 0.035 & 0.036 & 0.168 & 3 & 1.736 & 0.217 & 0.125 & 1.360 & 1.874\\
Bus-DeMeo Xc & 2 & 0.129 & 0.077 & 0.055 & 0.051 & 0.206 & 2 & 1.337 & 0.088 & 0.062 & 1.249 & 1.424\\
Bus-DeMeo Xe & 4 & 0.136 & 0.238 & 0.119 & 0.111 & 0.676 & 4 & 1.377 & 0.170 & 0.085 & 1.152 & 1.626\\
Bus-DeMeo Xk & 8 & 0.095 & 0.038 & 0.013 & 0.050 & 0.170 & 8 & 1.527 & 0.416 & 0.147 & 1.054 & 2.498\\
Tholen S & 502 & 0.210 & 0.084 & 0.004 & 0.037 & 0.830 & 465 & 1.598 & 0.449 & 0.021 & 0.467 & 3.591\\
Tholen C Complex & 406 & 0.057 & 0.072 & 0.004 & 0.020 & 0.769 & 358 & 1.065 & 0.405 & 0.021 & 0.124 & 3.934\\
Tholen C & 323 & 0.055 & 0.079 & 0.004 & 0.020 & 0.769 & 291 & 1.062 & 0.412 & 0.024 & 0.390 & 3.934\\
Tholen B & 52 & 0.082 & 0.035 & 0.005 & 0.034 & 0.204 & 36 & 0.904 & 0.308 & 0.051 & 0.563 & 1.674\\
Tholen F & 39 & 0.046 & 0.013 & 0.002 & 0.027 & 0.091 & 38 & 1.172 & 0.367 & 0.059 & 0.124 & 2.100\\
Tholen G & 12 & 0.067 & 0.040 & 0.011 & 0.035 & 0.200 & 12 & 1.032 & 0.840 & 0.242 & 0.390 & 3.814\\
Tholen V & 12 & 0.309 & 0.075 & 0.022 & 0.146 & 0.417 & 9 & 1.781 & 0.699 & 0.233 & 1.276 & 3.676\\
Tholen X Complex & 77 & 0.099 & 0.161 & 0.018 & 0.026 & 1.000 & 74 & 1.575 & 0.350 & 0.041 & 0.887 & 2.498\\
Tholen M & 33 & 0.125 & 0.037 & 0.006 & 0.064 & 0.224 & 33 & 1.623 & 0.291 & 0.051 & 1.108 & 2.498\\
Tholen E & 9 & 0.430 & 0.229 & 0.076 & 0.204 & 1.000 & 8 & 1.501 & 0.448 & 0.158 & 0.960 & 2.400\\
Tholen P & 35 & 0.044 & 0.014 & 0.002 & 0.026 & 0.112 & 33 & 1.511 & 0.375 & 0.065 & 0.887 & 2.423\\
Tholen Q & 1 & 0.165 & 0.000 & 0.000 & 0.165 & 0.165 & 1 & 1.897 & 0.000 & 0.000 & 1.897 & 1.897\\
Tholen D & 90 & 0.053 & 0.049 & 0.005 & 0.025 & 0.253 & 81 & 2.098 & 0.670 & 0.074 & 0.773 & 3.653\\
Tholen A & 27 & 0.224 & 0.076 & 0.015 & 0.110 & 0.410 & 26 & 1.746 & 0.568 & 0.111 & 0.926 & 3.244\\
Tholen R & 1 & 0.148 & 0.000 & 0.000 & 0.148 & 0.148 & 1 & 1.264 & 0.000 & 0.000 & 1.264 & 1.264\\
Tholen T & 34 & 0.094 & 0.067 & 0.011 & 0.036 & 0.413 & 30 & 1.529 & 0.389 & 0.071 & 0.762 & 2.384\\
\enddata
\end{deluxetable}

\begin{deluxetable}{llllllll}
\tabletypesize{\tiny}
\tablecolumns{8}
\tablecaption{Asteroids classified as X-types under either the Tholen, Bus, or Bus-DeMeo taxonomic schemes can be assigned Tholen-style M, E and P classes based on their visible albedos.  We assign the P type to objects with $p_{V}<0.1$, E to asteroids with $p_{V}>0.3$, and the rest to M type. The various X types are listed from the following sources: (1) \citet{Bus}, denoted Type 1; (2) \citet{Lazzaro}, denoted Type 2; (3) \citet{Lazzaro}, denoted Type 3; (4) \citet{Xu} (Type 4); (5) \citet{DeMeo09} (Type 5).}
\tablehead{
\colhead{Name} & \colhead{Type 1} & \colhead{Type 2} & \colhead{Type 3} & \colhead{Type 4} & \colhead{Type 5} & \colhead{$p_{V}$} & \colhead{New Tholen EMP Category} 
}
\startdata
22 & X &   &   &   & X & $ 0.168\pm0.038 $ & M\\
46 & Xc &   &   &   &   & $ 0.052\pm0.011 $ & P\\
56 & Xk & X & X &   & Xk & $ 0.050\pm0.010 $ & P\\
64 & Xe &   &   &   & Xe & $ 0.676\pm0.223 $ & E\\
71 & Xe &   &   &   &   & $ 0.247\pm0.051 $ & M\\
75 & Xk &   &   &   &   & $ 0.099\pm0.019 $ & P\\
76 & X &   &   &   & C & $ 0.049\pm0.010 $ & P\\
77 & Xe &   &   &   & Xe & $ 0.153\pm0.027 $ & M\\
83 & X &   &   &   &   & $ 0.086\pm0.021 $ & P\\
87 & X & X & X &   & X & $ 0.036\pm0.008 $ & P\\
97 &   &   &   &   & Xc & $ 0.206\pm0.046 $ & M\\
99 & Xk &   &   &   & Xk & $ 0.058\pm0.010 $ & P\\
107 & X & X & X &   &   & $ 0.055\pm0.013 $ & P\\
110 & X &   &   &   & Xk & $ 0.170\pm0.042 $ & M\\
114 & Xk &   &   &   & K & $ 0.088\pm0.010 $ & P\\
117 & X & X & X &   &   & $ 0.039\pm0.007 $ & P\\
125 & X &   &   &   &   & $ 0.115\pm0.022 $ & M\\
129 & X &   &   &   &   & $ 0.157\pm0.026 $ & M\\
131 & Xc &   &   & CX & K & $ 0.164\pm0.033 $ & M\\
132 & Xe &   &   &   & Xe & $ 0.119\pm0.022 $ & M\\
135 & Xk &   &   &   &   & $ 0.153\pm0.028 $ & M\\
136 & Xe &   &   &   &   & $ 0.164\pm0.033 $ & M\\
139 & X &   &   &   &   & $ 0.045\pm0.023 $ & P\\
143 & Xc &   &   &   &   & $ 0.053\pm0.011 $ & P\\
153 & X &   &   &   & X & $ 0.047\pm0.010 $ & P\\
164 & X & X & X &   &   & $ 0.043\pm0.007 $ & P\\
166 & Xe & Xk & X &   &   & $ 0.066\pm0.014 $ & P\\
181 & Xk & X & X &   & Xk & $ 0.079\pm0.015 $ & P\\
184 & X & X & X &   &   & $ 0.106\pm0.020 $ & M\\
190 & X &   &   &   &   & $ 0.038\pm0.008 $ & P\\
191 & Cb & X & X &   & Cb & $ 0.043\pm0.007 $ & P\\
199 & X & X & X &   & D & $ 0.116\pm0.026 $ & M\\
201 & X &   &   &   & Xk & $ 0.098\pm0.021 $ & P\\
209 & Xc &   &   &   &   & $ 0.058\pm0.010 $ & P\\
214 & Xc & B & B &   & Cg & $ 0.204\pm0.041 $ & M\\
216 & Xe &   &   &   & Xe & $ 0.111\pm0.034 $ & M\\
217 &   & X & X &   &   & $ 0.043\pm0.009 $ & P\\
220 &   & Xk & X &   &   & $ 0.057\pm0.011 $ & P\\
223 &   & Xc & X &   &   & $ 0.034\pm0.006 $ & P\\
224 &   & T & X &   &   & $ 0.161\pm0.031 $ & M\\
227 &   & X & X &   &   & $ 0.060\pm0.017 $ & P\\
231 &   &   &   & X &   & $ 0.066\pm0.014 $ & P\\
233 & K & T & T &   & Xk & $ 0.092\pm0.016 $ & P\\
242 & Xc &   &   &   &   & $ 0.160\pm0.027 $ & M\\
247 & Xc &   &   &   &   & $ 0.060\pm0.011 $ & P\\
248 &   &   &   & X &   & $ 0.048\pm0.019 $ & P\\
250 & Xk &   &   &   & Xk & $ 0.113\pm0.022 $ & M\\
255 &   & X & X &   &   & $ 0.033\pm0.008 $ & P\\
256 &   &   &   & X &   & $ 0.060\pm0.011 $ & P\\
259 & X & X & X &   &   & $ 0.042\pm0.009 $ & P\\
260 &   & X & X &   &   & $ 0.063\pm0.011 $ & P\\
261 & X &   &   &   &   & $ 0.101\pm0.015 $ & M\\
268 &   & X & X &   &   & $ 0.046\pm0.010 $ & P\\
272 & X &   &   &   &   & $ 0.127\pm0.018 $ & M\\
273 &   & Xk & K &   &   & $ 0.118\pm0.021 $ & M\\
279 & X &   &   &   & D & $ 0.039\pm0.006 $ & P\\
304 & Xc &   &   &   &   & $ 0.043\pm0.007 $ & P\\
307 &   & X & X &   &   & $ 0.040\pm0.011 $ & P\\
309 &   & X & X &   &   & $ 0.058\pm0.016 $ & P\\
317 & Xe &   &   &   &   & $ 0.505\pm0.056 $ & E\\
319 &   &   &   & X &   & $ 0.078\pm0.014 $ & P\\
322 & X &   &   &   & D & $ 0.074\pm0.008 $ & P\\
336 & Xk &   &   &   &   & $ 0.046\pm0.005 $ & P\\
338 & Xk &   &   &   &   & $ 0.163\pm0.032 $ & M\\
372 & B & X & C &   &   & $ 0.065\pm0.016 $ & P\\
373 &   & Ch & X &   &   & $ 0.047\pm0.011 $ & P\\
381 & Cb & X & C &   &   & $ 0.053\pm0.007 $ & P\\
388 & C & X & X &   &   & $ 0.044\pm0.010 $ & P\\
396 & Xe &   &   &   &   & $ 0.139\pm0.026 $ & M\\
409 & Xc &   &   &   &   & $ 0.050\pm0.009 $ & P\\
413 & X &   &   &   &   & $ 0.107\pm0.016 $ & M\\
415 &   & Xk & X &   &   & $ 0.086\pm0.016 $ & P\\
417 & Xk & X & X &   &   & $ 0.083\pm0.014 $ & P\\
418 &   & X & X &   &   & $ 0.106\pm0.018 $ & M\\
424 &   & Xc & X &   &   & $ 0.040\pm0.011 $ & P\\
426 &   & X & X &   &   & $ 0.056\pm0.009 $ & P\\
429 &   & Xk & X &   &   & $ 0.043\pm0.014 $ & P\\
436 &   & Xk & X &   &   & $ 0.046\pm0.008 $ & P\\
437 &   & Xc & X &   &   & $ 0.466\pm0.086 $ & E\\
441 & Xk &   &   &   &   & $ 0.139\pm0.024 $ & M\\
447 &   & X & X &   &   & $ 0.057\pm0.012 $ & P\\
455 &   & Xk & X &   &   & $ 0.045\pm0.008 $ & P\\
457 &   & Xk & X &   &   & $ 0.174\pm0.046 $ & M\\
461 &   & X & X &   &   & $ 0.048\pm0.008 $ & P\\
468 &   & Xk & X &   &   & $ 0.050\pm0.010 $ & P\\
469 &   & Xk & X &   &   & $ 0.043\pm0.012 $ & P\\
474 &   &   &   & X &   & $ 0.069\pm0.012 $ & P\\
491 & C & X & X &   &   & $ 0.051\pm0.010 $ & P\\
493 &   & X & X &   &   & $ 0.060\pm0.008 $ & P\\
504 & X & X & X &   &   & $ 0.251\pm0.040 $ & M\\
506 &   & X & X &   &   & $ 0.040\pm0.007 $ & P\\
507 & X &   &   &   &   & $ 0.133\pm0.026 $ & M\\
508 &   & X & X &   &   & $ 0.063\pm0.012 $ & P\\
511 & C & X & X &   &   & $ 0.071\pm0.011 $ & P\\
516 & X &   &   &   &   & $ 0.158\pm0.030 $ & M\\
522 &   & X & X &   &   & $ 0.057\pm0.013 $ & P\\
536 &   & X & X &   &   & $ 0.038\pm0.006 $ & P\\
543 & Xe &   &   &   &   & $ 0.152\pm0.020 $ & M\\
547 & Xk & T & T &   &   & $ 0.107\pm0.030 $ & M\\
558 &   & Xk & X &   &   & $ 0.120\pm0.018 $ & M\\
564 & Xc &   &   &   &   & $ 0.054\pm0.009 $ & P\\
567 &   & X & X &   &   & $ 0.053\pm0.006 $ & P\\
581 & Xk & X & X &   &   & $ 0.060\pm0.010 $ & P\\
589 &   & X & X &   &   & $ 0.040\pm0.008 $ & P\\
604 & Xc &   &   &   &   & $ 0.082\pm0.015 $ & P\\
607 &   & Ch & X &   &   & $ 0.040\pm0.007 $ & P\\
626 & Xc & Cb & C &   &   & $ 0.054\pm0.008 $ & P\\
627 & X &   &   &   &   & $ 0.094\pm0.016 $ & P\\
628 &   & Xc & X &   &   & $ 0.130\pm0.024 $ & M\\
629 & X &   &   &   &   & $ 0.089\pm0.017 $ & P\\
663 &   & X & X &   &   & $ 0.047\pm0.012 $ & P\\
671 & Xk &   &   &   &   & $ 0.046\pm0.015 $ & P\\
678 & X &   &   &   &   & $ 0.327\pm0.083 $ & E\\
680 &   & X & X &   &   & $ 0.046\pm0.007 $ & P\\
687 & X &   &   &   &   & $ 0.072\pm0.014 $ & P\\
696 &   & X & X &   &   & $ 0.056\pm0.011 $ & P\\
702 & B & X & C &   &   & $ 0.054\pm0.009 $ & P\\
705 & C & X & X &   &   & $ 0.046\pm0.010 $ & P\\
712 & X &   &   &   &   & $ 0.059\pm0.014 $ & P\\
713 & C & Ch & X &   &   & $ 0.043\pm0.008 $ & P\\
718 & X &   &   &   &   & $ 0.041\pm0.007 $ & P\\
731 & Xe &   &   &   &   & $ 0.257\pm0.051 $ & M\\
734 &   & X & X &   &   & $ 0.046\pm0.006 $ & P\\
739 & X & X & X &   & Xc & $ 0.051\pm0.012 $ & P\\
752 &   & Ch & Caa & X &   & $ 0.045\pm0.006 $ & P\\
757 & Xk &   &   &   &   & $ 0.110\pm0.015 $ & M\\
759 & X &   &   &   &   & $ 0.033\pm0.005 $ & P\\
768 &   & X & X &   &   & $ 0.141\pm0.029 $ & M\\
771 & X &   &   &   &   & $ 0.129\pm0.014 $ & M\\
779 & X & X & X &   &   & $ 0.174\pm0.056 $ & M\\
781 & Xc &   &   &   &   & $ 0.042\pm0.008 $ & P\\
789 & X &   &   &   & Xk & $ 0.139\pm0.027 $ & M\\
792 & X &   &   &   &   & $ 0.032\pm0.008 $ & P\\
796 & X & X & X &   &   & $ 0.205\pm0.041 $ & M\\
814 & C & X & X &   &   & $ 0.048\pm0.006 $ & P\\
816 &   & Xc & X &   &   & $ 0.044\pm0.008 $ & P\\
834 &   & X & X &   &   & $ 0.061\pm0.010 $ & P\\
844 & X &   &   &   &   & $ 0.126\pm0.022 $ & M\\
850 &   & X & X &   &   & $ 0.071\pm0.012 $ & P\\
859 &   & X & C &   &   & $ 0.060\pm0.011 $ & P\\
860 & X &   &   &   &   & $ 0.076\pm0.015 $ & P\\
866 & X &   &   &   &   & $ 0.041\pm0.008 $ & P\\
872 & X &   &   &   &   & $ 0.111\pm0.020 $ & M\\
882 &   & X & X &   &   & $ 0.064\pm0.009 $ & P\\
892 &   & X & X &   &   & $ 0.043\pm0.007 $ & P\\
894 &   & X & X &   &   & $ 0.115\pm0.022 $ & M\\
899 &   & X & X &   &   & $ 0.145\pm0.026 $ & M\\
907 & Xk &   &   &   &   & $ 0.027\pm0.007 $ & P\\
917 &   & X & X &   &   & $ 0.050\pm0.009 $ & P\\
928 &   & X & X &   &   & $ 0.038\pm0.007 $ & P\\
941 & X &   &   &   &   & $ 0.131\pm0.026 $ & M\\
943 &   & Ch & X &   &   & $ 0.047\pm0.007 $ & P\\
949 &   & Xk & X &   &   & $ 0.051\pm0.011 $ & P\\
952 &   & X & X &   &   & $ 0.047\pm0.004 $ & P\\
965 & Xc &   &   &   &   & $ 0.036\pm0.006 $ & P\\
972 &   & X & X &   &   & $ 0.037\pm0.005 $ & P\\
973 & Xk & X & X &   &   & $ 0.066\pm0.013 $ & P\\
977 &   & X & X &   &   & $ 0.054\pm0.009 $ & P\\
983 &   & Xk & X &   &   & $ 0.028\pm0.006 $ & P\\
1005 &   & Xk & X &   &   & $ 0.050\pm0.010 $ & P\\
1013 &   & Xk & X &   &   & $ 0.139\pm0.026 $ & M\\
1014 & Xe &   &   &   &   & $ 0.083\pm0.017 $ & P\\
1015 & Xc &   &   &   &   & $ 0.046\pm0.008 $ & P\\
1024 & Ch & X & Caa &   &   & $ 0.039\pm0.012 $ & P\\
1030 &   & X & X &   &   & $ 0.028\pm0.004 $ & P\\
1032 & X &   &   &   &   & $ 0.031\pm0.007 $ & P\\
1039 & X &   &   &   &   & $ 0.056\pm0.007 $ & P\\
1042 &   & X & Caa &   &   & $ 0.049\pm0.010 $ & P\\
1046 & Xe &   &   &   &   & $ 0.110\pm0.024 $ & M\\
1051 &   & Xc & X &   &   & $ 0.048\pm0.006 $ & P\\
1098 & Xe &   &   &   &   & $ 0.174\pm0.037 $ & M\\
1103 & Xk &   &   &   &   & $ 0.300\pm0.059 $ & E\\
1104 & Xk &   &   &   &   & $ 0.048\pm0.008 $ & P\\
1107 & Xc &   &   &   &   & $ 0.054\pm0.010 $ & P\\
1109 &   & X & D &   &   & $ 0.039\pm0.010 $ & P\\
1127 &   & X & X &   &   & $ 0.032\pm0.008 $ & P\\
1135 & Xk &   &   &   &   & $ 0.059\pm0.011 $ & P\\
1146 &   & X & X &   &   & $ 0.144\pm0.022 $ & M\\
1149 &   & X & X &   &   & $ 0.033\pm0.009 $ & P\\
1154 &   & X & X &   &   & $ 0.034\pm0.008 $ & P\\
1155 & Xe &   &   &   &   & $ 0.225\pm0.053 $ & M\\
1171 &   & X & X &   &   & $ 0.039\pm0.007 $ & P\\
1180 &   & Xe & X &   &   & $ 0.044\pm0.008 $ & P\\
1181 & X &   &   &   &   & $ 0.091\pm0.019 $ & P\\
1187 & X &   &   &   &   & $ 0.048\pm0.009 $ & P\\
1201 & Xc &   &   &   &   & $ 0.033\pm0.005 $ & P\\
1212 & X &   &   &   &   & $ 0.040\pm0.007 $ & P\\
1214 & Xk &   &   &   &   & $ 0.055\pm0.011 $ & P\\
1222 & X &   &   &   &   & $ 0.164\pm0.042 $ & M\\
1226 &   & Xk & D &   &   & $ 0.172\pm0.029 $ & M\\
1244 &   & X & X &   &   & $ 0.059\pm0.010 $ & P\\
1251 & X &   &   &   &   & $ 0.638\pm0.125 $ & E\\
1261 &   & X & X &   &   & $ 0.056\pm0.010 $ & P\\
1281 &   & X & X &   &   & $ 0.060\pm0.008 $ & P\\
1282 &   & Xe & X &   &   & $ 0.043\pm0.008 $ & P\\
1283 &   & X & X &   &   & $ 0.155\pm0.027 $ & M\\
1304 & X &   &   &   &   & $ 0.196\pm0.040 $ & M\\
1317 &   & Xk & X &   &   & $ 0.181\pm0.036 $ & M\\
1318 &   & Xe & X &   &   & $ 0.173\pm0.034 $ & M\\
1319 &   & X & X &   &   & $ 0.096\pm0.019 $ & P\\
1323 & Xc &   &   &   &   & $ 0.024\pm0.006 $ & P\\
1327 & X &   &   &   &   & $ 0.050\pm0.008 $ & P\\
1337 &   & Xk & X &   &   & $ 0.030\pm0.009 $ & P\\
1351 & Xk & Xc & X &   &   & $ 0.067\pm0.013 $ & P\\
1352 & X &   &   &   &   & $ 0.145\pm0.019 $ & M\\
1355 &   & Xe & X &   &   & $ 0.467\pm0.114 $ & E\\
1356 &   & X & X &   &   & $ 0.054\pm0.011 $ & P\\
1373 & Xk &   &   &   &   & $ 0.152\pm0.024 $ & M\\
1420 & X &   &   &   &   & $ 0.096\pm0.018 $ & P\\
1424 & X &   &   &   &   & $ 0.062\pm0.011 $ & P\\
1428 & Xc &   &   &   &   & $ 0.025\pm0.008 $ & P\\
1436 &   & X & X &   &   & $ 0.033\pm0.005 $ & P\\
1463 &   &   &   & X &   & $ 0.071\pm0.015 $ & P\\
1469 &   & X & X &   &   & $ 0.074\pm0.014 $ & P\\
1490 & Xc &   &   &   &   & $ 0.104\pm0.024 $ & M\\
1493 & Xc &   &   &   &   & $ 0.069\pm0.010 $ & P\\
1517 & X &   &   &   &   & $ 0.039\pm0.006 $ & P\\
1541 & Xc &   &   &   &   & $ 0.097\pm0.019 $ & P\\
1546 &   & X & X &   &   & $ 0.115\pm0.016 $ & M\\
1548 & Xk &   &   &   &   & $ 0.045\pm0.008 $ & P\\
1571 &   & Xc & X &   &   & $ 0.128\pm0.020 $ & M\\
1585 &   & X & X &   &   & $ 0.029\pm0.006 $ & P\\
1592 & X &   &   &   &   & $ 0.220\pm0.039 $ & M\\
1605 &   & X & X &   &   & $ 0.187\pm0.034 $ & M\\
1628 &   &   &   & X &   & $ 0.049\pm0.007 $ & P\\
1638 & X &   &   &   &   & $ 0.117\pm0.018 $ & M\\
1653 & X &   &   & C &   & $ 0.668\pm0.117 $ & E\\
1693 &   & X & X &   &   & $ 0.047\pm0.008 $ & P\\
1712 &   &   &   & X &   & $ 0.050\pm0.010 $ & P\\
1730 & Xe &   &   &   &   & $ 0.189\pm0.035 $ & M\\
1765 &   & X & X &   &   & $ 0.136\pm0.025 $ & M\\
1796 & Cb & X & X &   &   & $ 0.044\pm0.008 $ & P\\
1819 &   & X & X &   &   & $ 0.058\pm0.009 $ & P\\
1841 &   & X & X &   &   & $ 0.057\pm0.010 $ & P\\
1847 & Xc &   &   &   &   & $ 0.231\pm0.040 $ & M\\
1860 & X &   &   &   &   & $ 0.100\pm0.015 $ & P\\
1919 &   & Xe & X &   &   & $ 0.701\pm0.034 $ & E\\
1936 & Ch & X & X &   &   & $ 0.057\pm0.004 $ & P\\
1992 &   & Xk & X &   &   & $ 0.145\pm0.031 $ & M\\
1995 &   &   &   & X &   & $ 0.063\pm0.051 $ & P\\
1998 & Xc &   &   &   &   & $ 0.107\pm0.021 $ & M\\
2001 & Xe & Xe & X &   &   & $ 0.841\pm0.145 $ & E\\
2065 & Xc &   &   &   &   & $ 0.084\pm0.013 $ & P\\
2073 & X &   &   &   &   & $ 0.154\pm0.030 $ & M\\
2103 &   & X & X &   &   & $ 0.139\pm0.021 $ & M\\
2104 &   & X & X &   &   & $ 0.104\pm0.019 $ & M\\
2140 &   &   &   & X &   & $ 0.053\pm0.007 $ & P\\
2194 & Xc &   &   &   &   & $ 0.183\pm0.031 $ & M\\
2204 &   & X & X & X &   & $ 0.050\pm0.006 $ & P\\
2303 &   & X & X &   &   & $ 0.295\pm0.058 $ & M\\
2306 & X &   &   &   &   & $ 0.132\pm0.014 $ & M\\
2349 & Xc & Xk & X &   &   & $ 0.166\pm0.031 $ & M\\
2390 & X &   &   &   &   & $ 0.042\pm0.007 $ & P\\
2407 &   & X & X &   &   & $ 0.150\pm0.029 $ & M\\
2444 & C &   &   & X &   & $ 0.053\pm0.007 $ & P\\
2489 &   & X & Caa &   &   & $ 0.059\pm0.009 $ & P\\
2491 &   & Xe & X &   &   & $ 0.544\pm0.102 $ & E\\
2507 & Xe &   &   &   &   & $ 0.133\pm0.022 $ & M\\
2559 & Xk &   &   &   &   & $ 0.049\pm0.006 $ & P\\
2560 & Xc &   &   &   &   & $ 0.102\pm0.014 $ & M\\
2567 & Xc &   &   &   &   & $ 0.156\pm0.024 $ & M\\
2606 & Xk &   &   &   &   & $ 0.176\pm0.031 $ & M\\
2634 &   & X & X &   &   & $ 0.108\pm0.021 $ & M\\
2681 & Xk &   &   &   &   & $ 0.228\pm0.090 $ & M\\
2736 & Xc &   &   &   &   & $ 0.848\pm0.236 $ & E\\
2861 & Xc &   &   &   &   & $ 0.069\pm0.011 $ & P\\
2879 & X &   &   &   &   & $ 0.067\pm0.013 $ & P\\
2996 & Xc &   &   &   &   & $ 0.069\pm0.012 $ & P\\
3007 & X &   &   &   &   & $ 0.147\pm0.024 $ & M\\
3109 &   &   &   & X &   & $ 0.064\pm0.017 $ & P\\
3169 & Xe & Cb & C &   &   & $ 0.413\pm0.095 $ & E\\
3256 & X &   &   &   &   & $ 0.047\pm0.007 $ & P\\
3262 & X &   &   &   &   & $ 0.138\pm0.025 $ & M\\
3328 &   & Xc & K &   &   & $ 0.148\pm0.030 $ & M\\
3330 &   & X & X &   &   & $ 0.048\pm0.008 $ & P\\
3367 & X &   &   &   &   & $ 0.303\pm0.059 $ & E\\
3381 &   &   &   & X &   & $ 0.517\pm0.124 $ & E\\
3406 & X &   &   &   &   & $ 0.158\pm0.025 $ & M\\
3440 & X &   &   &   &   & $ 0.174\pm0.030 $ & M\\
3445 &   & X & X &   &   & $ 0.055\pm0.007 $ & P\\
3451 & X &   &   &   &   & $ 0.049\pm0.012 $ & P\\
3483 &   & Xk & X &   &   & $ 0.862\pm0.088 $ & E\\
3567 & Xc &   &   &   &   & $ 0.087\pm0.017 $ & P\\
3575 & X &   &   &   &   & $ 0.201\pm0.039 $ & M\\
3615 &   & X & C &   &   & $ 0.086\pm0.016 $ & P\\
3670 & X &   &   &   &   & $ 0.064\pm0.013 $ & P\\
3686 & X &   &   &   &   & $ 0.064\pm0.011 $ & P\\
3691 & Xc &   &   &   &   & $ 0.672\pm0.158 $ & E\\
3704 & Xk &   &   &   &   & $ 0.181\pm0.035 $ & M\\
3740 &   &   &   & X &   & $ 0.071\pm0.012 $ & P\\
3762 & X &   &   &   &   & $ 0.513\pm0.113 $ & E\\
3789 &   & Xk & T &   &   & $ 0.099\pm0.016 $ & P\\
3832 &   & X & C &   &   & $ 0.069\pm0.016 $ & P\\
3865 & Xc &   &   &   &   & $ 0.238\pm0.041 $ & M\\
3880 &   & Xe & X &   &   & $ 0.574\pm0.130 $ & E\\
3915 &   & Xc & C & C &   & $ 0.049\pm0.005 $ & P\\
3939 &   & X & X &   &   & $ 0.042\pm0.009 $ & P\\
3940 &   & T & X &   &   & $ 0.641\pm0.108 $ & E\\
3958 & Xc &   &   &   &   & $ 0.574\pm0.085 $ & E\\
3976 & X &   &   &   &   & $ 0.038\pm0.010 $ & P\\
3985 & X &   &   &   &   & $ 0.152\pm0.027 $ & M\\
4006 &   &   &   & X &   & $ 0.070\pm0.002 $ & P\\
4031 &   &   &   & X &   & $ 0.398\pm0.092 $ & E\\
4165 &   &   &   & XS &   & $ 0.123\pm0.025 $ & M\\
4201 &   & X & X &   &   & $ 0.061\pm0.013 $ & P\\
4256 & Xc &   &   &   &   & $ 0.210\pm0.024 $ & M\\
4342 & Xc &   &   &   &   & $ 0.068\pm0.010 $ & P\\
4353 & Xe &   &   & X &   & $ 0.138\pm0.024 $ & M\\
4369 & Xk &   &   &   &   & $ 0.120\pm0.024 $ & M\\
4424 & Xk &   &   &   &   & $ 0.073\pm0.014 $ & P\\
4440 &   &   &   & X &   & $ 0.567\pm0.033 $ & E\\
4460 &   & X & X &   &   & $ 0.041\pm0.008 $ & P\\
4461 & X &   &   &   &   & $ 0.135\pm0.025 $ & M\\
4483 &   & X & X &   &   & $ 0.215\pm0.038 $ & M\\
4547 & X &   &   &   &   & $ 0.039\pm0.007 $ & P\\
4548 & Xc &   &   &   &   & $ 0.206\pm0.042 $ & M\\
4613 &   & Xe & S &   &   & $ 0.284\pm0.036 $ & M\\
4701 & Xe &   &   &   &   & $ 0.053\pm0.005 $ & P\\
4750 & X &   &   &   &   & $ 0.087\pm0.010 $ & P\\
4764 &   & X & X &   &   & $ 0.896\pm0.118 $ & E\\
4786 & Xc &   &   &   &   & $ 0.534\pm0.104 $ & E\\
4838 & Xc &   &   &   &   & $ 0.105\pm0.020 $ & M\\
4839 & Xc &   &   &   &   & $ 0.204\pm0.039 $ & M\\
4845 & X &   &   &   &   & $ 0.181\pm0.018 $ & M\\
4942 & X &   &   &   &   & $ 0.631\pm0.135 $ & E\\
4956 &   &   &   & XT &   & $ 0.167\pm0.034 $ & M\\
5087 & X &   &   &   &   & $ 0.064\pm0.007 $ & P\\
5294 & X &   &   &   &   & $ 0.175\pm0.042 $ & M\\
5301 &   & X & C &   &   & $ 0.070\pm0.012 $ & P\\
5343 &   & X & X &   &   & $ 0.276\pm0.042 $ & M\\
5467 & X &   &   &   &   & $ 0.115\pm0.023 $ & M\\
5588 & X &   &   &   &   & $ 0.163\pm0.031 $ & M\\
5632 & Xc &   &   &   &   & $ 0.192\pm0.036 $ & M\\
6051 &   & X & X &   &   & $ 0.324\pm0.044 $ & E\\
6057 &   & X & X &   &   & $ 0.043\pm0.011 $ & P\\
6249 & Xe &   &   &   &   & $ 0.786\pm0.147 $ & E\\
6394 &   & Xe & X &   &   & $ 0.637\pm0.131 $ & E\\
8795 &   & X & C &   &   & $ 0.136\pm0.018 $ & M\\
10261 &   & Xk & X &   &   & $ 0.079\pm0.004 $ & P\\
11785 & Xc &   &   &   &   & $ 0.101\pm0.021 $ & M\\
12281 & X &   &   &   &   & $ 0.040\pm0.006 $ & P\\
\enddata
\end{deluxetable}

\begin{figure}
\figurenum{1}
\includegraphics[width=6in]{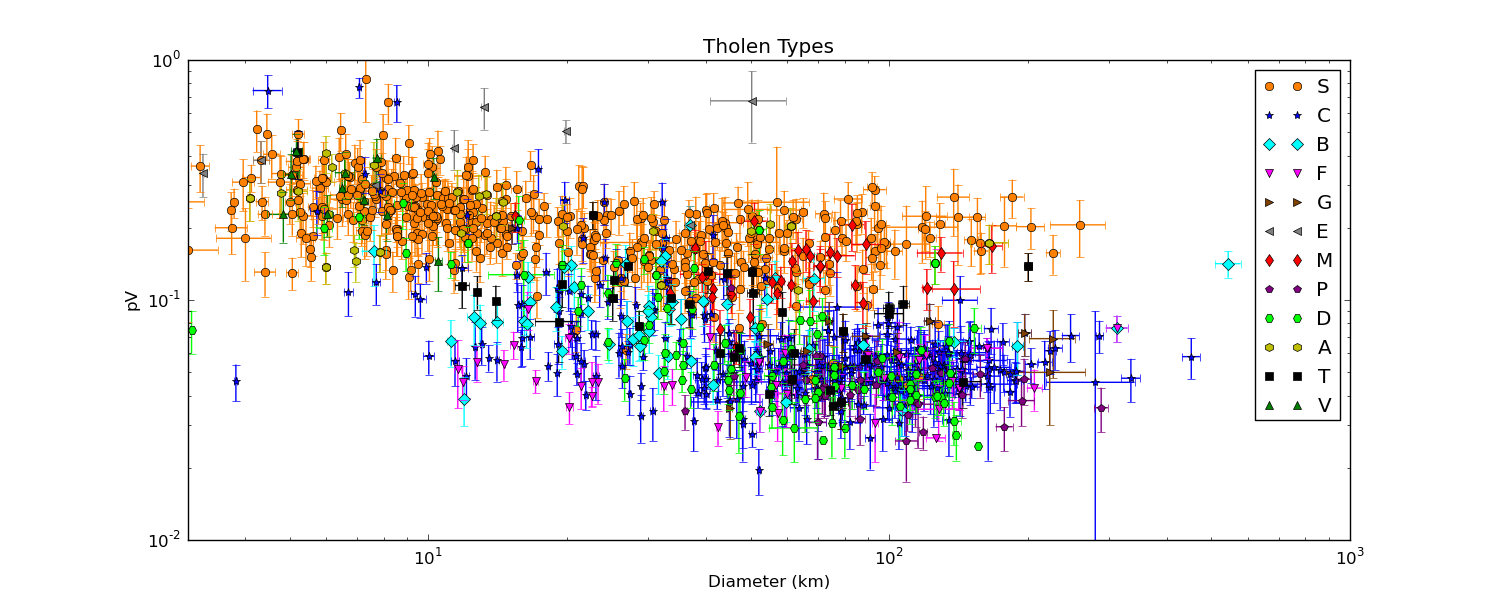}
\caption{\label{fig:tholen_diam_alb} NEOWISE-derived diameters vs. albedos of asteroids observed and classified according to the Tholen system.  The Tholen system preserves the albedo distinctions between its different spectral classes very well down to $\sim$30 km, at which point selection biases begin to become apparent in that low albedo objects are missing. Furthermore, this bias is likely to be at least partially, if not entirely, responsible for the apparent increase in albedo with decreasing diameter for all taxonomic types. }
\end{figure}

\clearpage

\end{document}